\documentclass[reprint,superscriptaddress,amsmath,amssymb,aps,pra]{revtex4-2}

\usepackage[utf8]{inputenc}
\usepackage[T1]{fontenc}
\usepackage{graphicx}
\usepackage{dcolumn}
\usepackage{bm}
\usepackage{color}
\usepackage{hyperref}
\usepackage{upgreek,textgreek}
\usepackage{physics}
\usepackage{orcidlink}

\newcommand{\Hop}{\hat{H}}
\newcommand{\aop}{\hat{a}}
\newcommand{\aopd}{\hat{a}^\dag}
\newcommand{\nop}{\hat{n}}

\newcommand{\Wnet}{\mathcal{W}_{\text{net}}}
\newcommand{\W}[2]{W_{\text{#1}\to\text{#2}}}
\newcommand{\Q}[2]{Q_{\text{#1}\to\text{#2}}}
\newcommand{\U}[2]{\mathcal{U}_{\text{#1}\to\text{#2}}}
\newcommand{\DS}[2]{\Delta S_{\text{#1}\to\text{#2}}}
\newcommand{\COPR}{\text{COP}_{\text{R}}}
\newcommand{\etac}{\eta^{(\text{Carnot})}}

\begin{document}

\title{A Bose-Hubbard quantum thermal machine}

\author{Felipe Isaule~\orcidlink{0000-0003-1810-0707}}
\email{felipe.isaule@usm.cl}
\affiliation{Departamento de Física, Universidad Técnica Federico Santa María, Av. España 1680, Valparaíso 2390123, Chile}

\author{Juan E. Bernate}
\affiliation{Quantum Systems Unit, Okinawa Institute of Science and Technology Graduate University, Onna, Okinawa 904-0495, Japan}
\affiliation{Department of Physics and Arnold Sommerfeld Center for Theoretical Physics (ASC), Ludwig-Maximilians-Universität München, Theresienstr. 37, München D-80333, Germany}

\author{Francisco J. Peña~\orcidlink{0000-0002-7432-0707}}
\affiliation{Facultad de Ingeniería, Universidad San Sebastián, Lago Panguipulli 1390, Puerto Montt, Chile}

\author{Patricio Vargas~\orcidlink{0000-0001-9235-9747}}
\affiliation{Departamento de Física, Universidad Técnica Federico Santa María, Av. España 1680, Valparaíso 2390123, Chile}

\author{Thomás Fogarty~\orcidlink{0000-0003-4940-5861}}
\affiliation{Quantum Systems Unit, Okinawa Institute of Science and Technology Graduate University, Onna, Okinawa 904-0495, Japan}

\date{\today}

\begin{abstract}

This work proposes a quantum thermal machine comprised of interacting ultracold bosons confined in a one-dimensional optical lattice and modeled by the Bose-Hubbard Hamiltonian. We consider an Otto-inspired cycle with strokes that change the interaction strength between bosons, in addition to standard thermalization and compression/expansion strokes. We find that these interaction changes can be manipulated to enhance the performance of the thermal machine, outperforming non-interacting cycles. In particular, when operating as a heat engine, the cycle can achieve its largest work output and near-Carnot efficiency by compressing the medium at strong repulsion into a Mott insulator state. Additionally, we find that decreasing interactions after the compression stroke favors and enhances the engine operation, whereas increasing interactions favors refrigeration cycles. The findings of this work further show that controlling the interactions across quantum phases of matter can be harnessed as a useful thermodynamic resource. 

\end{abstract}

\maketitle

\section{Introduction}

Quantum thermal machines (QTMs) have emerged as a fundamental platform for studying quantum thermodynamics~\cite{kosloff_quantum_2014,vinjanampathy_quantum_2016,myers_quantum_2022,cangemi_quantum_2024,potts_quantum_2026}. They enable us to examine the role of quantum effects in thermodynamic devices and to assess whether they offer advantages over their classical counterparts. Over the years, a variety of quantum resources have been considered, including phase transitions~\cite{campisi_power_2016,ma_quantum_2017,chand_critical-point_2018,b_s_universal_2020,myers_boosting_2022}, quantum statistics~\cite{zheng_quantum_2015,myers_bosons_2020,watanabe_quantum_2020,myers_thermodynamics_2021,dunlop_thermodynamics_2026}, and squeezed reservoirs~\cite{huang_effects_2012,rosnagel_nanoscale_2014,manzano_entropy_2016,klaers_squeezed_2017}. Within such resources, collective quantum many-body effects have received increased attention, as they can further boost the performance of QTMs~\cite{jaramillo_quantum_2016,niedenzu_cooperative_2018} and enable us to harness the role of interactions~\cite{chen_bose-fermi_2018,wang_performance_2020,s_quantum_2025}.

One ideal testbed for studying many-body effects in quantum thermodynamics is ultracold atomic gases. They offer a highly controllable setting~\cite{bloch_many-body_2008,bloch_quantum_2012}, leading to landmark experimental realizations of quantum Otto cycles~\cite{bouton_quantum_2021,nettersheim_power_2022}. One of the celebrated properties of ultracold atoms is the ability to manipulate the inter-atomic interactions through Feshbach resonance techniques~\cite{chin_feshbach_2010}. This controllability has motivated proposals for QTMs that change the strength of interactions across the thermodynamic cycle~\cite{chen_interaction-driven_2019,keller_feshbach_2020,boubakour_interaction-enhanced_2023,nautiyal_finite-time_2024,watson_quantum_2025,wandhammer_optimal_2026}, as well as recent experimental realizations of some of these ideas~\cite{simmons_thermodynamic_2023,koch_quantum_2023}. In particular, Ref.~\cite{koch_quantum_2023} realized the Pauli engine, which drove a Fermi gas across the BCS-BEC crossover by tuning the interactions, changing the statistics of the working medium within a single cycle. Such statistical changes were later expanded in Refs.~\cite{menon_leveraging_2025,henriquez_lira_quantum_2026} within hybrid Otto cycles. Thus, ultracold atoms offer a controllable setting to harness interactions as a quantum resource, which can be combined with statistical and phase-transition effects.

While most studies of ultracold QTMs have focused on configurations in free space, ultracold atoms can also be confined in optical lattices, offering additional scenarios for quantum simulation~\cite{bloch_ultracold_2005,lewenstein_ultracold_2012}. Indeed, ultracold atoms in optical lattices exhibit a plethora of quantum phases, such as insulators and crystalline states, that could be harnessed as quantum resources. In this direction, Ref.~\cite{fogarty_many-body_2021} proposed a lattice engine that drives interacting bosons between superfluid (SF) and Mott insulator (MI) states by varying the lattice depth from shallow to deep regimes. In contrast, non-interacting bosons were considered in Ref.~\cite{pena_bosonic_2026}, where a cycle is driven by a controllable flux in a Rhombic chain. Additionally, Ref.~\cite{lal_bera_hybrid_2026} manipulated the atom statistics on a lattice in the Anyon Hubbard model, showing that changes in statistics can boost performance. However, the manipulation of interactions in optical lattices as a thermodynamic quantum resource remains largely unexplored.

In this work, we propose a QTM composed of interacting bosons in a one-dimensional tight optical lattice, which we model with the Bose-Hubbard Hamiltonian~\cite{fisher_boson_1989,jaksch_cold_1998}. We focus on the manipulation of interactions as a quantum resource by considering an Otto-inspired cycle with strokes that change the interatomic interaction strength. By complementing these strokes with changes in the lattice size, we can exploit the thermodynamic response associated with entering and leaving the MI regime. As a proof of principle, this work considers a minimal engine composed of a few bosons, which can be realized in current ultracold atom laboratories~\cite{di_carli_commensurate_2024}. We find that our thermodynamic cycle can operate as an engine for a wide range of parameters. In particular, the engine can deliver its maximum work output and very high efficiency when the interaction strength is reduced after the working medium is compressed into a strongly repulsive MI state. In turn, the cycle behaves as a high-performance refrigerator when the interactions are instead increased after the compression stroke.

This work is organized as follows. Firstly, Sec.~\ref{sec:model} presents the theoretical model and the numerical approach. Sec.~\ref{sec:cycle} then presents the thermodynamic cycle, detailing each stroke and how the work and heat exchanges are calculated. The main results of this work are presented in Sec.~\ref{sec:results}, including an exhaustive examination of the performance of the engine, an analysis of the available thermodynamic operations, and a brief examination of the performance of the refrigeration cycle. Finally, Sec.~\ref{sec:concl} summarizes the main conclusions of this work. Complementary results are also reported in the appendix.

\section{Model}
\label{sec:model}

\begin{figure*}[t!]
\centering
\includegraphics[width=\textwidth]{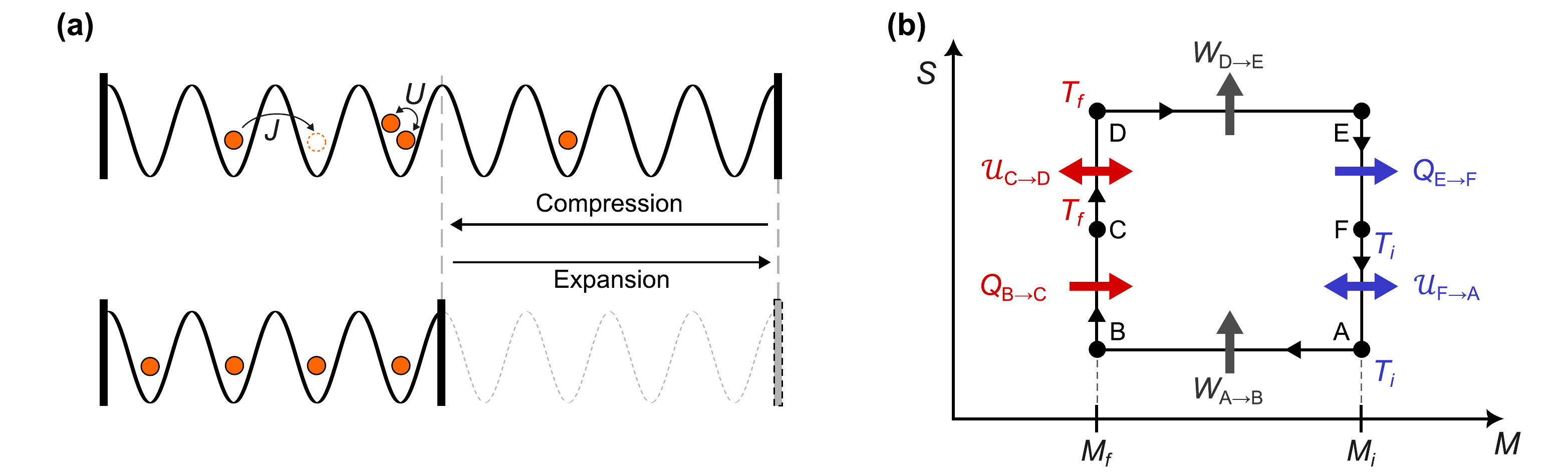}
\caption{(a) Illustration of the system under consideration. The bosons (orange circles) interact via on-site repulsive interactions and can tunnel to neighboring sites. The horizontal arrows depict the compression and expansion of the lattice. (b) Illustration of the cycle in terms of the entropy $S$ and the number of sites $M$. The figure depicts a typical engine operation with $T_f>T_i$.
}
\label{sec:model;fig:illustration}
\end{figure*}

We consider a working medium composed of a fixed number of $N$ bosons that interact via short-range repulsive interactions. The bosons are confined in an open one-dimensional optical lattice, where the number of sites $M$ changes across the compression and expansion strokes. We depict this configuration in Fig.~\ref{sec:model;fig:illustration}(a). In the tight-binding regime, the system at a given state is described by the Bose-Hubbard Hamiltonian~\cite{fisher_boson_1989,jaksch_cold_1998}

\begin{equation}
    \Hop = - J\sum_{i=1}^{M-1}\left(\aopd_{i}\aop_{i+1}+\textrm{h.c.}\right)+\frac{U}{2}\sum_{i=1}^M\nop_{i}(\nop_i-1),
    \label{sec:model;eq:H}
\end{equation}
where $\aop_{i}$ and $\aopd_{i}$ annihilate and create a boson at site $i$, respectively. In turn, $\nop_{i}=\aopd_{i}\aop_{i}$ is the number operator. The first term describes the tunneling with amplitude $J>0$, while the second term describes the on-site boson-boson interaction with strength $U>0$.

We assume that in a given state, the system is in thermal equilibrium with a bath at temperature $T$. The thermodynamics are then extracted from the canonical partition function $Z=\sum_n e^{-\beta E_n}$, where $E_n$ are the eigenenergies and $\beta=1/(k_B T)$ is the inverse temperature. Thus, expectation values are obtained from $\langle \hat{O} \rangle = Z^{-1}\sum_n \langle \Psi_n|\hat{O}|\Psi_n\rangle e^{-\beta E_n}$, with $|\Psi_n\rangle$ the eigenvectors. Throughout this work, we calculate differences in the internal energy between states $a$ and $b$ from $\U{a}{b}=\langle \Hop_b\rangle-\langle \Hop_a\rangle$, where $\langle \Hop \rangle = Z^{-1}\sum_n E_n e^{-\beta E_n}$. Additionally, the entropy is calculated from $S=k_{B}(\log Z+\beta \langle \Hop \rangle)$, which is used to find isentropic strokes $\Delta S_{a\to b}=S_b-S_a=0$.

We compute the full energy spectrum within the exact diagonalization (ED) method for a fixed number of particles~\cite{zhang_exact_2010,raventos_cold_2017}. This allows us to compute the partition function $Z$ exactly. While ED restricts the calculations to small systems, it enables us to fully calculate excited states and to accurately compute thermodynamic observables across a wide range of temperatures. The latter is crucial for accurately performing isentropic strokes.

Nevertheless, our proposed setup with a few bosons and sites has been realized experimentally in recent years~\cite{di_carli_commensurate_2024}. Such work also implemented lattice size changes through controllable deep potential barriers, which make selected edge sites energetically inaccessible. Importantly, because the number of atoms remains constant, the physical full Hilbert space does not change. Therefore, our effective model with varying $M$ is valid at temperatures where the high-energy states of the potential barrier are not accessed.

Before presenting the cycle, we mention that our objective is to leverage the different physical states of our working medium. At low temperatures ($k_B T\lesssim J$), our model~(\ref{sec:model;eq:H}) displays either a Mott insulator (MI) or, for the few bosons considered here, a superfluid (SF) state. Importantly, the MI state appears only for large repulsion ($U/J\gtrsim 4$) and commensurate filling~\cite{raventos_cold_2017,kuhner_phases_1998}, as illustrated by the bottom illustration of Fig.~\ref{sec:model;fig:illustration}(a). In contrast, at high temperatures ($k_B T\gtrsim J$) the system behaves as a normal (N) thermal fluid. These states will be accessed through compression, expansion, and thermalization strokes. However, we stress that our small finite system displays crossovers instead of well-defined phase transitions. Nevertheless, our results focus on parameters deep in each of the regimes so we can isolate their effect. We refer to App.~\ref{app:thermo} for additional details on these physical properties.

\section{Cycle}
\label{sec:cycle}

\begin{table}[b!]
\caption{\label{sec:cycle;tab:States} Temperature $T$, number of sites $M$, and interaction $U$ at each of the six states within the cycle. The temperatures $T'_i$ and $T'_f$ are obtained by imposing isentropic strokes $S_B=S_A$ and $S_E=S_D$, respectively.}
\begin{ruledtabular}
\begin{tabular}{lccc}
State & $T$ & $M$ & $U$  \\
\hline
A       & $T_i$ & $M_i$ & $U_i$  \\
B       & $T'_i$ & $M_f$ & $U_i$  \\
C       & $T_f$ & $M_f$ & $U_i$  \\
D       & $T_f$ & $M_f$ & $U_f$  \\
E       & $T'_f$ & $M_i$ & $U_f$  \\
F       & $T_i$ & $M_i$ & $U_f$  \\
\end{tabular}
\end{ruledtabular}
\end{table}

We consider an Otto-inspired cycle which operates between two thermal reservoirs at temperatures $T_i$ and $T_f$, between two lattice sizes $M_i$ and $M_f$, and between two interaction strengths $U_i$ and $U_f$. Without loss of generality, the tunneling coupling $J$ is kept fixed. Similar to the engine studied in Ref.~\cite{menon_leveraging_2025}, our cycle consists of six strokes across the six states summarized in Table~\ref{sec:cycle;tab:States}. The cycle is depicted in Fig.~\ref{sec:model;fig:illustration}(b). 

The strokes are as follows: 
\begin{enumerate}
    \item \emph{Adiabatic compression} (A$\to$B): Initially, the system is in thermal equilibrium with the reservoir at temperature $T_i$. Additionally, the lattice is initially composed of $M_i$ lattice sites, and the bosons interact with strength $U_i$. After decoupling the reservoir, the system is adiabatically ($\DS{A}{B}=0$) compressed from $M_i$ to $M_f<M_i$ sites, and thus the temperature is allowed to change from $T_i$ to $T'_i$. The interaction is kept fixed. Because this stroke is adiabatic, only work is done, which is calculated from $\W{A}{B} =\langle \Hop_{\text{B}}  \rangle-\langle \Hop_{\text{A}} \rangle$.
    \item \emph{First isochoric thermalization} (B$\to$C): While maintaining a fixed number of sites $M_f$ and a fixed interaction coupling $U_i$, the system is coupled to the reservoir at temperature $T_f$. Therefore, the temperature changes from $T'_i$ to $T_f$. During this stroke, only heat is exchanged, which is calculated from $\Q{B}{C} =\langle \Hop_{\text{C}}  \rangle-\langle \Hop_{\text{B}} \rangle$.
    \item \emph{First isothermal interaction change} (C$\to$D): With the system coupled to the final reservoir, the interaction strength is changed from $U_i$ to $U_f$. In turn, the temperature and lattice size are kept fixed at $T_f$ and $M_f$, respectively. During this stroke, both heat and work are exchanged, and thus the change in internal energy reads $\U{C}{D}=\W{C}{D}+\Q{C}{D}$. Because this stroke is isothermal, the heat is calculated directly from the entropy difference through $\Q{C}{D}=T_f(S_D-S_C)$. Then, the work is calculated from $\W{C}{D} =\langle \Hop_{\text{D}} \rangle-\langle \Hop_{\text{C}} \rangle-\Q{C}{D}$.
    \item \emph{Adiabatic expansion} (D$\to$E): After decoupling the reservoir,  the system is adiabatically ($\DS{D}{E}=0$) expanded back from $M_f$ to $M_i$ sites, while the temperature changes from $T_f$ to $T'_f$. The interaction is kept fixed. In this adiabatic stroke, the work is calculated from $\W{D}{E} =\langle \Hop_{\text{E}}  \rangle-\langle \Hop_{\text{D}} \rangle$.
    \item \emph{Second isochoric thermalization} (E$\to$F): While maintaining a fixed number of sites $M_i$ and a fixed interaction coupling $U_f$, the system is coupled again to the initial reservoir. Thus, the temperature changes back from $T'_f$ to $T_i$. The exchanged heat is calculated from $\Q{E}{F} =\langle \Hop_{\text{F}}  \rangle-\langle \Hop_{\text{E}} \rangle$.
    \item \emph{Second isothermal interaction change} (F$\to$A): Finally, with the system still coupled to the initial reservoir, the interaction is changed back from $U_f$ to $U_i$. The temperature and lattice sizes are kept at the initial values $T_i$ and $M_i$, respectively. As with the previous isothermal stroke, both heat and work are exchanged $\U{F}{A}=\W{F}{A}+\Q{F}{A}$. Therefore, heat is calculated from $\Q{F}{A}=T_i(S_A-S_F)$, while the work is then calculated from $\W{F}{A} =\langle \Hop_{\text{A}} \rangle-\langle \Hop_{\text{F}} \rangle-\Q{F}{A}$.
\end{enumerate}

From the previous definitions, work and heat are negative if they exit the system, and positive if they enter. Thus, the net work output of the cycle reads
\begin{equation}
    \Wnet=-(\W{A}{B}+\W{C}{D}+\W{D}{E}+\W{F}{A}).
    \label{sec:cycle;eq:Wnet}
\end{equation}
In turn, for a standard cycle with reservoirs such that $T_i<T_f$, the heat exchanges are defined as
\begin{equation}
    Q_c=\Q{E}{F}+\Q{F}{A},\qquad Q_h=\Q{B}{C}+\Q{C}{D},
    \label{sec:cycle;eq:Q}
\end{equation}
corresponding to the heat exchanges with the cold ($c$) and hot ($h$) reservoirs. Thus, in these cases the first thermalization corresponds to a heating stroke, while the second thermalization corresponds to a cooling stroke. While most of the results presented throughout this work consider $T_i<T_f$, we will also briefly examine the inverse operation where $T_i>T_f$. In the latter, the heat exchanges are defined inversely, $Q_h=\Q{E}{F}+\Q{F}{A}$ and $Q_c=\Q{B}{C}+\Q{C}{D}$.

It is important to note that the compression (A$\to$B) and expansion (D$\to$E) strokes are modeled within a quasi-static equilibrium approximation. The working medium is assumed to remain in internal thermal equilibrium while thermally isolated from the reservoirs, with no entropy production. These conditions define the idealized thermodynamic description adopted here. Thus, the microscopic dynamics required to realize such transformations are beyond the scope of this work.

To simplify the following discussion, we focus on three distinct protocols for the interactions. Firstly, we consider a fixed interaction strength throughout the cycle $U_i=U_f=U$. This corresponds to the limit of the standard four-stroke Otto cycle. Secondly, we consider cycles that increase the interaction from the non-interacting limit $U_i=0$ to $U_f=U$. Thirdly, we consider cycles that instead decrease the interaction from $U_i=U$ to $U_f=0$. Nevertheless, more general interaction changes are examined in App.~\ref{app:interactions}.

\section{Results}
\label{sec:results}

\begin{figure*}[t!]
\centering
\includegraphics[width=\textwidth]{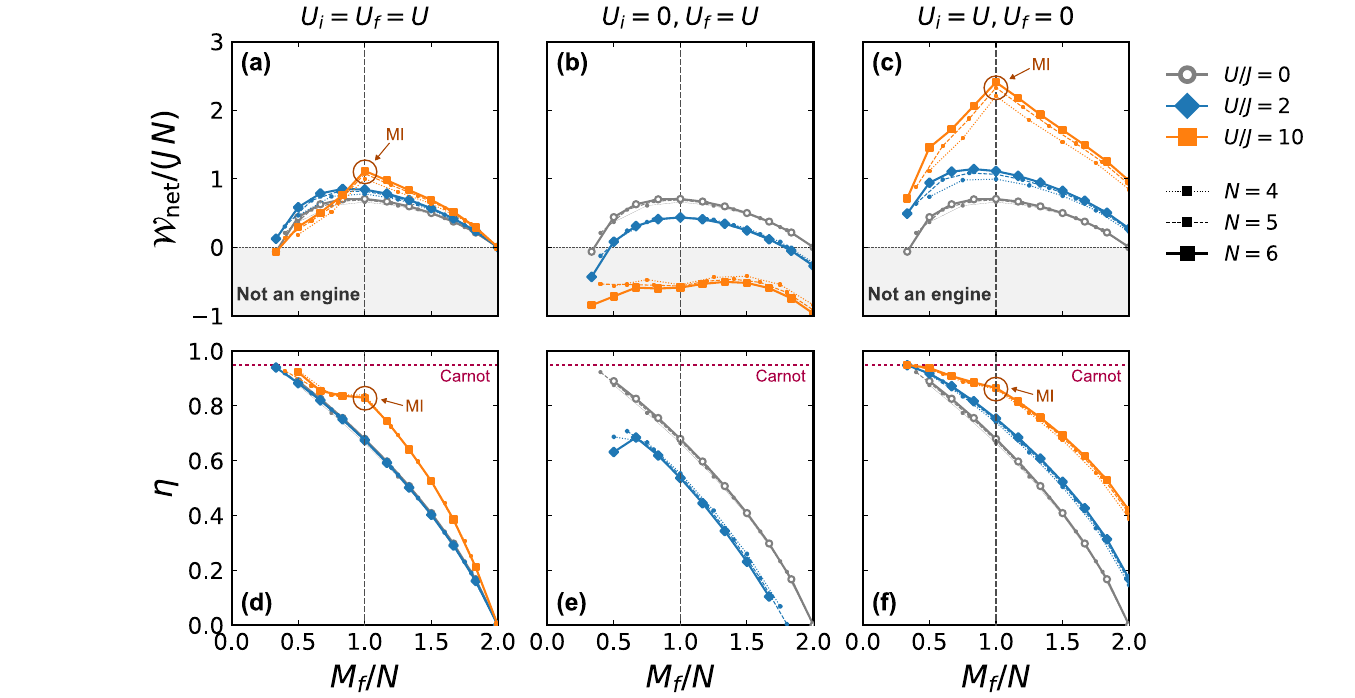}
\caption{Net work output per particle (a--c) and engine efficiency (d--e) as a function of the final lattice size $M_f$. The calculations consider an initial lattice size of $M_i/N=2$ and the protocol of interactions indicated above the panels. The engine operates between reservoirs with temperatures $k_B T_i/J=0.2$ and $k_B T_f/J=4.0$. Each curve considers the interaction strengths and number of particles indicated in the legend. The gray region in (a--c) indicates that the thermal machine does not operate as an engine, while the horizontal dotted line in (d--f) indicates the Carnot efficiency. In panel (d), the non-interacting results are almost completely underneath the results for $U/J=2$.}
\label{sec:results;sub:engine;fig:M}
\end{figure*}

\subsection{Engine performance}
\label{sec:results;sub:engine}

We start by examining the engine operation of our thermal machine. An engine converts heat from a hot source to useful work. Thus, an engine operation is defined by a positive net work output $\Wnet>0$, and heat exchanges such that $Q_c<0$ and $Q_h>0$. The efficiency of the engine is defined as
\begin{equation}
    \eta=\frac{\Wnet}{Q_h},
\end{equation}
which takes values $0\leq \eta \leq 1$. Here, $\eta$ is to be compared with the Carnot efficiency $\etac=1-T_c/T_h$, where $T_c$ and $T_h$ are the temperatures of the cold and hot reservoirs, respectively.

\subsubsection{Dependence on the compression}

Firstly, to examine how the performance depends on the compression, we show in Fig.~\ref{sec:results;sub:engine;fig:M} the net work output and efficiency as a function of the final lattice size $M_f$. We note that here and in the following, we use a lattice with an initial size of $M_i/N=2$, corresponding to half-filling. 
We compare the non-interacting limit (gray markers) with weak (blue markers) and strong (orange markers) interactions. Note that the three protocols are equivalent for $U=0$. Additionally, the figure displays results for different numbers of particles, showing a very weak dependence on $N$. Thus, our results are robust for the small systems considered in this work.

For the fixed interaction protocol (left panels), one immediately notes that the work output [panel (a)] shows a maximum at a finite value of $M_f$ for the three choices of $U$. For weak interactions in the SF regime ($U\lesssim 4$), the work output shows a weak dependence on $U$, behaving similarly to non-interacting Otto cycles. In contrast, we observe a notable peak at exactly unit filling $M_f/N=1$ for large repulsion (orange markers), as highlighted by the orange circle. This corresponds to the emergence of the MI state. Therefore, the largest work output is obtained when compressing the engine to an MI state across stroke A$\to$B. These results are in line with the findings from Ref.~\cite{fogarty_many-body_2021}, where the largest work output is obtained when inducing an MI state through potential depth changes. 

The efficiency [panel (d)] does increase with the compression (decreasing $M_f$), approaching the Carnot efficiency (dashed horizontal line) for small $M_f$. However, this has the drawback of a vanishing work output [see panel (a)]. Interestingly, the efficiency is approximately identical between no and weak interactions. In contrast, larger efficiencies can be reached over almost the entire range of $M_f$ when considering strong repulsive interactions $U/J=10$. Importantly, at unit filling, the efficiency displays a peak (highlighted by the orange circle), corresponding to the maximum work output in panel (a). Therefore, we can conclude that the best performance, meaning both large work output and efficiency, is obtained by driving the engine through an MI state by using a large $U$ and compressing down to $M_f/N=1$. These results already show that the interplay between interactions and lattice effects can enhance the engine's performance.

For the increasing interaction protocol (middle panels), we observe that the cycle behaves as an engine only for weak interactions. Indeed, for strong interactions (orange markers), we obtain a negative work output for any compression [panel (b)]. Nevertheless, the performance of the cycle shows a similar dependence on $M_f$ for vanishing and small $U$ (gray and blue markers). However, the largest net work and efficiency are now associated with the non-interacting limit. Therefore, these results suggest that increasing the interactions through the cycle significantly hinders the engine's performance.

Finally, for the decreasing interaction protocol (right panels), the cycle behaves as an engine in all the shown cases. As in panel (a), the work output for strong repulsion reaches its maximum at unit filling $M_f/N=1$, as highlighted by the orange circle. Crucially, at such a maximum, the work output reaches a value of $\Wnet/(JN)\approx 2.5$. This is approximately three times larger than the values obtained for weak interactions and for the fixed interaction protocol [panel (a)]. In turn, the efficiency [panel (f)] increases with decreasing $M_f$, and shows larger values for strong interactions. Therefore, these results suggest that the best performance is obtained when compressing the medium at strong repulsion to an MI state (stroke A$\to$B), and later decreasing the inter-atomic interactions (stroke C$\to$D). Our next step is then to further examine whether the engine's performance shows a systematic behavior in terms of $U$ and to understand the physical origin of these performance differences.

 Before continuing, however, it is also interesting to examine the behavior at $M_f=M_i$ ($M_f/N=2$ in the figure). This corresponds to the limit of a four-stroke cycle with no compression/expansion. For the Otto cycle with fixed interactions [panel (a)], the work output is simply zero, as there is no work done during the entire cycle. On the other hand, when changing the interactions across the cycle [panels (b) and (c)], the work output is finite at $M_f/N=2$. While counterintuitive, this means that our thermal machines can exchange work purely from interaction changes, with no need for traditional compression and expansion processes. This highlights the crucial role that interactions play in the improved performance of these hybrid Otto cycles. Nevertheless, the peak performance of the cycle is consistently obtained when compressing the system to the MI regime. Therefore, in the following, we focus on analyzing the engine when compressing to unit filling $M_f/N=1$.

\subsubsection{Dependence on the interaction}

\begin{figure*}[t!]
\centering
\includegraphics[width=\textwidth]{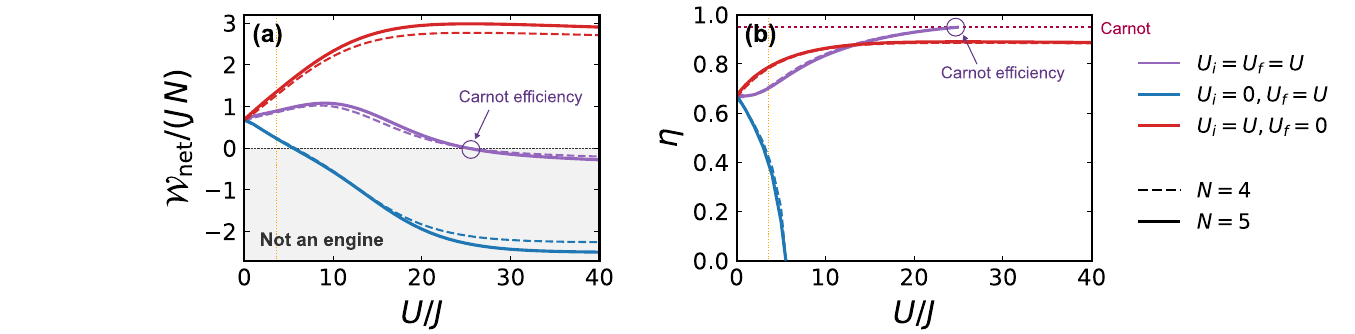}
\caption{Net work output per particle (a) and engine efficiency (b) as a function of the protocol's interaction $U$. 
The calculations consider an initial lattice size of $M_i/N=2$ and a final lattice size of $M_f/N=1$. The engine operates between reservoirs with temperatures $k_B T_i/J=0.2$ and $k_B T_f/J=4.0$. Each curve considers the protocol of interactions and number of particles indicated in the legend. The gray region in (a) indicates that the thermal machine does not operate as an engine, while the horizontal dotted line in (b) indicates the Carnot efficiency. The vertical orange dotted lines in both panels indicate the SF–MI phase transition interaction in the thermodynamic limit $U/J\approx 3.6$~\cite{kuhner_phases_1998}.}
\label{sec:results;sub:engine;fig:U}
\end{figure*}

We now examine how the performance depends on the interaction strength to better understand its role within the cycle. Thus, we show the net work output and efficiency as a function of $U$ for the three protocols in Fig.~\ref{sec:results;sub:engine;fig:U}. As previously mentioned, we focus on the case with $M_f/N=1$.

For equal interactions (purple curves), the work output [panel (a)] initially increases with $U$, reaching a maximum approximately at $U/J\approx 10$. Then, it decreases, vanishing at $U/J\approx 25$. Conversely, the efficiency [panel (b)] increases with $U$ within the engine's operating regime. In particular, $\eta$ reaches the Carnot efficiency at the point where $\Wnet=0$ (indicated by the purple circle), as expected for an Otto cycle. Overall, we can conclude that, for this protocol, the best performance is obtained at approximately $U/J\approx 10$, that is, with state B deep in the MI state. Thus, the compression to an MI state provides a performance increase over the non-interacting Otto cycle.

Then, for the increasing interaction protocol (blue curves), we observe only a small region where the engine operates. Indeed, the work output is only positive for weak interactions, $U/J\lesssim 5$. In contrast, the cycle does not operate as an engine when the MI state is accessed. Furthermore, both the work output and efficiency decrease for increasing $U$, showing the largest performance at $U=0$. Therefore, increasing the interactions after the compression stroke worsens the performance, confirming our previous conclusions.

Finally, for the decreasing interaction protocol (red curves), the cycle operates as an engine for all the examined interactions. Importantly, this figure clearly illustrates that this protocol offers the largest performance, with a work output of $\Wnet/(JN)>2$ for $U/J\gtrsim 10$. This large work output surpasses the other protocols and provides a noticeable advantage over the non-interacting medium. Moreover, for the chosen temperatures, the efficiency quickly saturates to $\eta\approx 0.85$, close to the Carnot efficiency. 

In summary, the presented results show that the best performance is obtained when performing the compression at strong repulsion. However, one might naively expect the opposite. Because the energy of the system increases with $U$, compressing at strong repulsion (stroke A$\to$B) requires a large work input, which could decrease the engine's performance. Thus, in the following we examine the thermodynamics of the system to understand what is driving the engine's performance.

\begin{figure*}[t!]
\centering
\includegraphics[width=\textwidth]{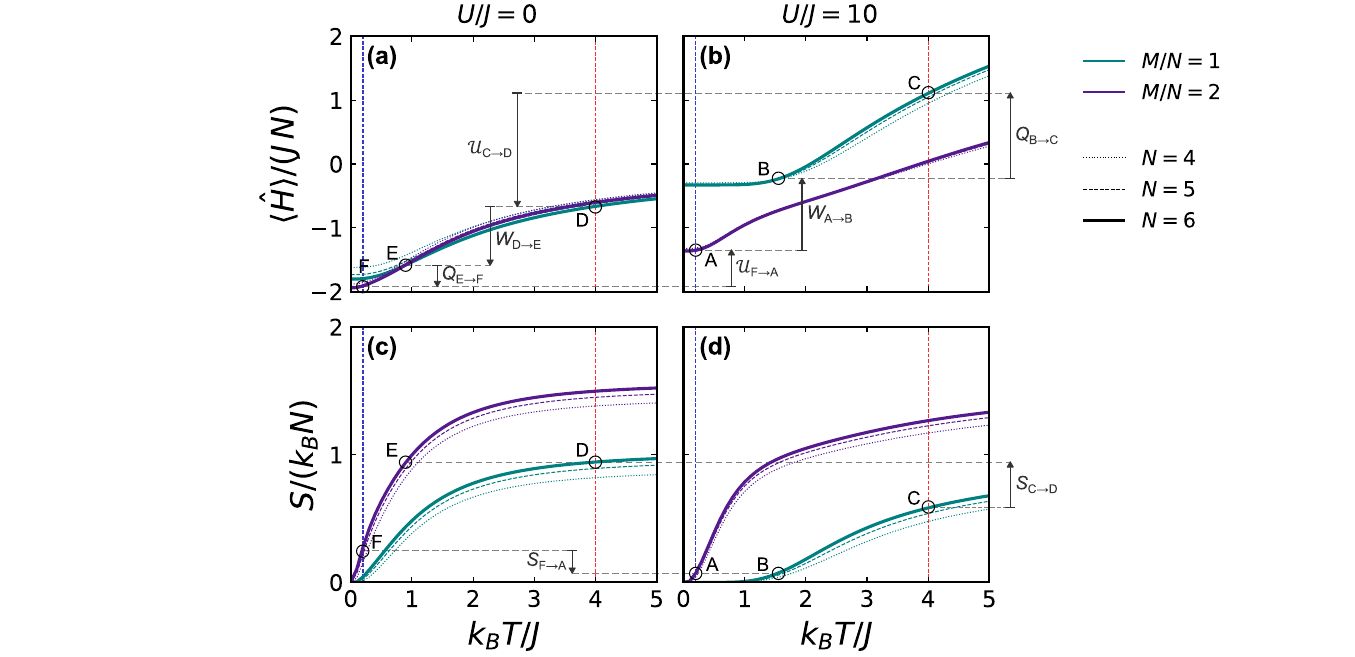}
\caption{Energy [(a) and (b)] and entropy [(c) and (d)] per particle of the working medium as a function of temperature. Each curve shows the lattice size and the number of particles indicated in the legend, while each column of panels shows the interaction strengths indicated above the figure. The vertical dashed blue lines on the left of each panel indicate $k_B T/J=0.2$, while the vertical dashed red lines on the right indicate $k_BT/J =4.0$. The circles indicate the states in a cycle with the decreasing interaction protocol for $k_B T_i/J=0.2$ and $k_B T_f/J=4.0$, corresponding to the cycle highlighted with the orange circles in Fig.~\ref{sec:results;sub:engine;fig:M}(c) and (f). The energy and entropy exchanges between such states are indicated by the arrows.}
\label{sec:results;sub:engine;fig:thermo}
\end{figure*}

\subsubsection{Thermodynamics}

To illustrate the thermodynamics driving the engine's performance, we show representative examples of the energy and entropy as a function of temperature in Fig.~\ref{sec:results;sub:engine;fig:thermo}. We show these quantities at $U/J=0$ and $U/J=10$, as well as for unit ($M/N=1$) and half ($M/N=2$) filling to compare with the previously reported results. The figure displays results for different particle numbers, further showing that the thermodynamics are similar for the small systems considered in this work.

Firstly, we note that the energy is not overly sensitive to the lattice size in the non-interacting limit [panel (a)], as all the curves overlap. In contrast, we observe larger energy differences between unit and half filling with strong interactions [panel (b)], which stem from the large energy gaps in the MI regime. This energy difference is harnessed by the fixed and decreasing interaction protocols to reach a large energy at state C. In turn, the entropy (bottom panels) is larger in panel (c) than in panel (d), showing that $S$ increases with decreasing $U$. This increase is also harnessed by the decreasing interaction protocol to absorb additional heat when the interaction strength is decreased.  In contrast, within the increasing interaction protocol, the lower energy increase in panel (a) means that not much heat is absorbed during the heating stroke. Consequently, the lower entropy in panel (b) means that a large amount of heat is lost when increasing the interactions at a high temperature, which leads to its ineffectiveness as an engine.

To better illustrate how work and heat are exchanged in each stroke, we now qualitatively examine the energy and entropy differences within a representative cycle with the decreasing interaction protocol (see circles in Fig.~\ref{sec:results;sub:engine;fig:thermo}). Across the compression A$\to $B, we observe a somewhat large work input of $\W{A}{B}/(JN)\approx 1.1$ due to the MI state. As previously mentioned, this is the inherently large energy cost that is required to compress the strongly interacting system, which could hinder the performance of the engine. However, the MI state also allows the system to later absorb a large amount of heat from the hot bath. Indeed, due to the rapid increase in energy with temperature at strong repulsion, a significant amount of heat is absorbed during the first thermalization B$\to$C ($\Q{B}{C}/(JN)\approx 1.4$ in the reference cycle). This heat is later converted to useful work by both the fixed and decreasing interaction protocols at large $U$.

Afterwards, additional heat is absorbed when decreasing interactions during stroke C$\to$D. Due to the increase in entropy with decreasing $U$, we observe that $\DS{C}{D}/(k_B N)\approx 0.4$ (see bottom panels), resulting in a total heat input of $\Q{C}{D}/(JN)\approx 1.6$, as the medium is coupled to the hot reservoir. In turn, because the energy decreases strongly at unit filling with decreasing $U$, we observe a significant energy decrease of $\U{C}{D}/(JN)\approx -1.8$. Overall, this combination of energy and entropy results in a large work output of $\W{C}{D}/(JN)\approx -3.4$, which is responsible for the enhanced performance of the decreasing interaction protocol. Additionally, a moderate amount of work $\W{D}{E}/(JN)\approx -0.9$ is later released during the expansion stroke. At this point, the system reaches a low energy and entropy, and thus not much heat can be lost in the following thermal strokes. Indeed, the second thermalization shows a small heat output of $\Q{E}{F}/(JN)\approx -0.4$. In turn, across the stroke F$\to$A we observe a very small entropy difference $\DS{F}{A}/(k_B N)\approx -0.2$, resulting in an almost negligible heat output $\Q{F}{A}$ because of the small value of $T_i$. Combining the latter with the small energy difference of $\U{F}{A}/(JN)\approx 0.6$ results in a final small work input of $\W{F}{A}/(JN)\approx 0.6$. Overall, these energy and entropy differences produce the large net work output of $\Wnet/(JN)\approx 2.5$ reported in Fig.~\ref{sec:results;sub:engine;fig:M}(c), which far exceeds the other cycles.

\begin{figure*}[t!]
\centering
\includegraphics[width=\textwidth]{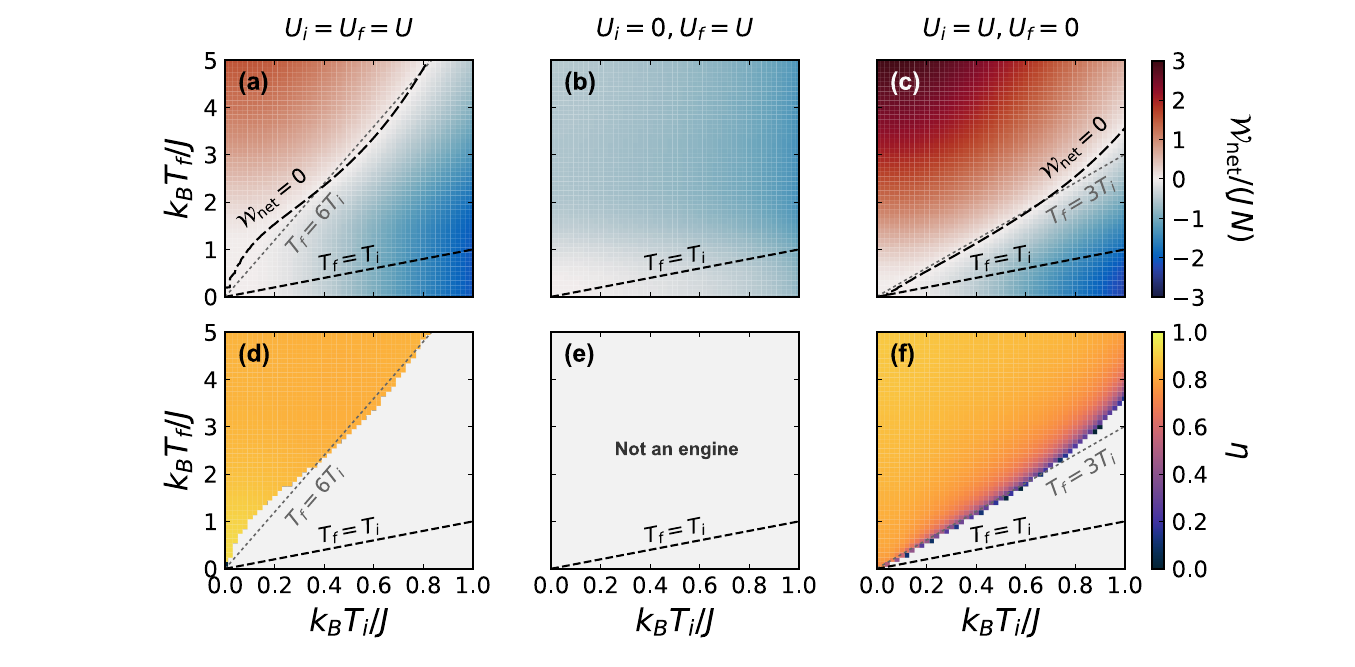}
\caption{Net work output per particle (a--c) and engine efficiency (d--f) as a function of the temperatures $T_i$ and $T_f$. 
The calculations consider $N=6$, a protocol interaction of $U/J=10$, an initial lattice size of $M_i/N=2$, and a final lattice size of $M_f/N=1$. The protocol of interactions is indicated above the panels. The long-dashed black curves in (a) and (c) indicate a vanishing work output $\Wnet=0$, while the gray regions in the bottom panels indicate that the thermal machine does not operate as an engine.  The diagonal dashed black line in each panel indicates $T_f=T_i$, while the dotted gray lines in the left panels and in the right panels indicate $T_f=6T_i$ and $T_f=3T_i$, respectively.
}
\label{sec:results;sub:engine;fig:T}
\end{figure*}

\subsubsection{Dependence on the temperatures}

The previous analysis assumed a large ratio $T_f/T_i$, so the absorbed heat surpasses the work required during the compression. Thus, we conclude our study of the engine operation by examining its dependence on the two bath temperatures. We show the engine's performance as a function of both $T_i$ and $T_f$ in Fig.~\ref{sec:results;sub:engine;fig:T} with compressions to unit filling and at strong repulsion. For completeness, the figure displays results for reversed reservoirs such that $T_i>T_f$ (below the dashed black line). 

For fixed interactions (left panels), the cycle operates as an engine for a large ratio $T_f/T_i$ (upper-left region of the panel), confirming our previous claim. In particular, for the chosen interaction strength, this ratio approximately fulfills $T_f\gtrsim 6T_i$ (gray dotted line). 
Furthermore, within this operational region, the work output [panel (a)] increases with increasing $T_f/T_i$, reaching a value of $\Wnet/(JN)\approx 1.5$ in the shown region. In contrast, the efficiency [panel (d)] is approximately constant. While the latter means that the efficiency remains large for a wide range of temperatures, the ratio $\eta/\etac$ decreases with increasing $T_f$. Here we stress that this constant efficiency is because both the work and heat exchanges display a similar dependence on $T_i$ and $T_f$. Interestingly, a similar behavior was reported in the anyon-Hubbard engine of Ref.~\cite{lal_bera_hybrid_2026}.

Then, for the increasing interactions protocol (middle panels), the cycle does not operate as an engine for any of the temperatures examined, as $\Wnet<0$ in the whole region [panel (b)]. Overall, this further confirms that this protocol is not useful for engine operations, as $\Wnet>0$ only for weak interactions, where the performance is nevertheless hindered.

Lastly, for the decreasing interactions protocol (right panels), the cycle also operates as an engine for $T_f/T_i>1$.
Particularly, for the chosen interaction strength, we found that it operates as an engine for approximately $T_f\gtrsim 3T_i$ (gray dotted line).
Importantly, this protocol displays a much larger operational window than for fixed interactions because of the heat absorbed during stroke C$\to$D.
Additionally, the work output [panel (c)] is enhanced, as $\Wnet$ reaches values up to $\Wnet/(JN)\approx 3$ (upper-left region of the panel), surpassing by a factor of two the fixed interaction protocol. In turn, the efficiency [panel (f)] is also large and approximately constant within the shown region. These results further confirm that the decreasing interaction protocol offers the best enhancement of the engine's performance across a wide range of temperatures.

\begin{figure*}[t!]
\centering
\includegraphics[width=\textwidth]{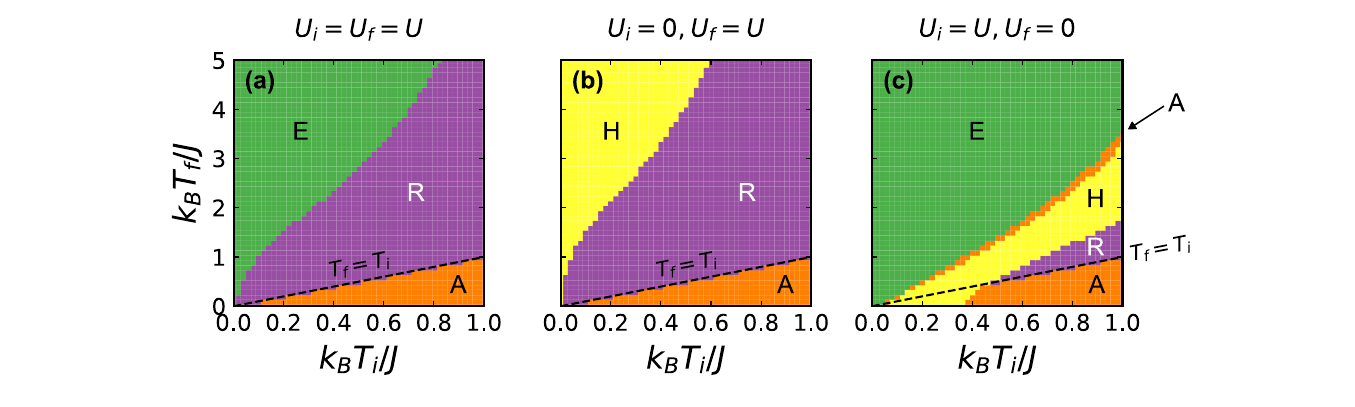}
\caption{Thermal operations as a function of the temperatures $T_i$ and $T_f$. The green region indicates an engine (E), the purple region indicates a refrigerator (R), the yellow region indicates a heater (H), and the orange region indicates an accelerator (A).
The calculations consider $N=6$, a protocol interaction of $U/J=10$, an initial lattice size of $M_i/N=2$, and a final lattice size of $M_f/N=1$. The protocol of interactions is indicated above the panels. The dashed black line in each panel indicates $T_f=T_i$. }
\label{sec:results;sub:operations;fig:diagram}
\end{figure*}

To finish our analysis, we stress again that the temperature $T_f$ must be large enough to excite the system in the compression branch so a large amount of heat can be absorbed from the hot reservoir. In particular, because the energy gap in the MI regime is $\Delta E\approx U-4J$, the critical ratio $T_f/T_i$ for engine operation increases with $U$, which we have checked numerically. This raises a trade-off, as while a small $U$ increases the temperature window of operation, it decreases the engine's performance because the MI regime might not be accessed. Thus, we have concluded that an interaction strength of $U/J\approx 10$ gives a near-optimal performance with a low critical ratio of $T_f/T_i<10$ for both the fixed and decreasing interaction protocols.

\subsection{Operations}
\label{sec:results;sub:operations}

So far, we have demonstrated that interactions can be manipulated to enhance the engine's performance. However, the interactions can also be manipulated to control the net work input ($\Wnet<0$), as shown in the upper panels of Fig.~\ref{sec:results;sub:engine;fig:T}. This work input could then be used for other operations, as studied with related thermal machines~\cite{menon_leveraging_2025,watson_quantum_2025,lal_bera_hybrid_2026}.

Depending on the work output and heat exchanges, the operational modes of our thermal machine can be classified as
\begin{itemize}
    \item Engine: $\Wnet \geq 0$, $Q_h\geq 0$, $Q_c\leq 0$.
    \item Refrigerator: $\Wnet \leq 0$, $Q_h\leq 0$, $Q_c\geq 0$.
    \item Accelerator: $\Wnet \leq 0$, $Q_h\geq 0$, $Q_c\leq 0$.
    \item Heater: $\Wnet \leq 0$, $Q_h\leq 0$, $Q_c\leq 0$.
\end{itemize}
In a refrigerator, work is needed to further cool a cold reservoir, and thus, $\Wnet<0$ and $Q_c \geq 0$. In contrast, an accelerator uses work to accelerate the heat transfer from the hot to the cold reservoir. Finally, a heater uses work to heat both reservoirs.

A representative diagram with the operational modes as a function of the temperatures is shown in Fig.~\ref{sec:results;sub:operations;fig:diagram}. This uses the same parameters as in Fig.~\ref{sec:results;sub:engine;fig:T}. Thus, the cycle operates as an engine (green region) only at large $T_f/T_i$ for fixed [panel (a)] and decreasing [panel (c)] interactions, corresponding to the region where $\Wnet>0$. In turn, for $\Wnet<0$, the cycle can behave as different operations depending on the parameters.

For fixed interactions [panel (a)] and $\Wnet<0$, the cycle behaves only as a refrigerator (purple region) for $T_f>T_i$. Thus, to get a refrigerator, the temperature ratio $T_f/T_i$ must be greater than one, but not large enough to operate as an engine. In turn, the cycle simply behaves as an accelerator (orange region) for $T_f<T_i$. Interestingly, this operational diagram is in qualitative agreement with the related Otto-like cycle of Ref.~\cite{menon_leveraging_2025}, which considers a one-dimensional gas in free space.

For the increasing interactions protocol [panel (b)], where $\Wnet<0$ in the whole plotted region, the cycle behaves again as a refrigerator for moderate temperature ratios $T_f/T_i>1$. However, it behaves as a heater (yellow region) for larger $T_f/T_i$, almost mirroring the engine region in panel (a). Finally, the cycle again behaves as an accelerator for $T_f<T_i$.

Finally, for the decreasing interactions protocol [panel (c)], we observe a much richer behavior for $\Wnet<0$. For $T_f>T_i$, the cycle behaves as a refrigerator only in a small region with $T_f\gtrsim T_i$. At slightly larger $T_f/T_i$, we observe instead a well-defined region with a heater operation, and then a narrow accelerator band at $\Wnet \lesssim 0$. On the other hand, for $T_f<T_i$, the cycle displays a heater at small temperatures $k_B T_i/J\lesssim 0.3$, and an accelerator for higher temperatures $T_i$.

Overall, the operational diagrams suggest that the increasing interactions protocol favors refrigerator operations, while decreasing interactions favor engine operations. To examine this further, we briefly examine the performance of the refrigeration cycle.

\begin{figure*}[t!]
\centering
\includegraphics[width=\textwidth]{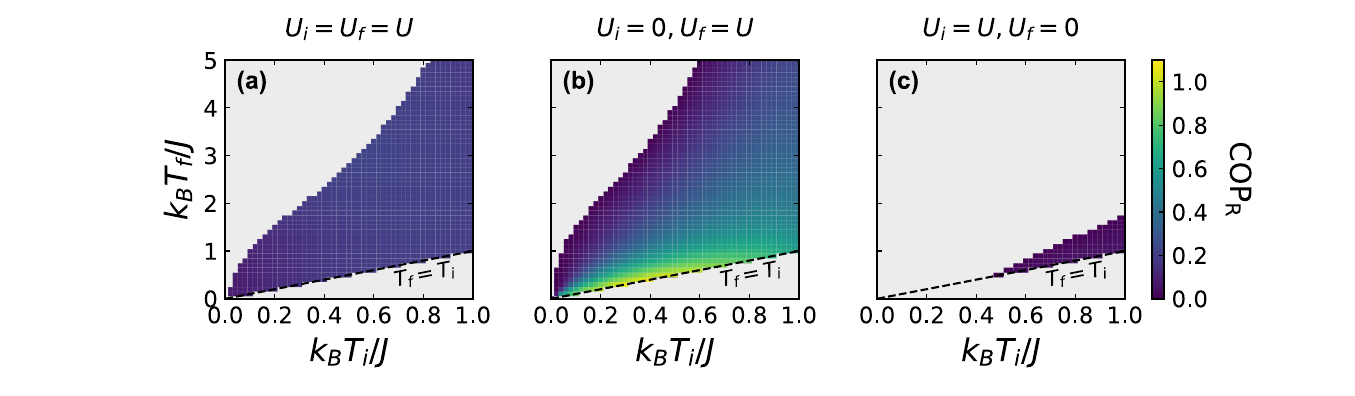}
\caption{Coefficient of performance of the refrigeration cycle as a function of the temperatures $T_i$ and $T_f$. 
The calculations consider $N=6$, a protocol interaction of $U/J=10$, an initial lattice size of $M_i/N=2$, and a final lattice size of $M_f/N=1$. The interaction protocol is indicated above the panels. The gray regions indicate that the thermal machine does not operate as a refrigerator. The diagonal dashed black line in each panel indicates $T_f=T_i$.}
\label{sec:results;sub:refrigerator;fig:T}
\end{figure*}

\subsection{Refrigeration cycle}
\label{sec:results;sub:refrigerator}

The performance of the refrigeration cycle is characterized by the coefficient of performance $\COPR$.
This is defined as
\begin{equation}
    \COPR=\frac{Q_c}{|\Wnet|},
\end{equation}
which is always positive, as $Q_c>0$. However, unlike the efficiency, $\COPR$ is not bounded by one.

In the following, we compare $\COPR$ with the Carnot coefficient of performance $\COPR^{\text{(Carnot)}}=(T_h/T_c-1)^{-1}$, with $T_h$ and $T_c$ the temperatures of the hot and cold reservoirs, respectively. Importantly, $\COPR^{\text{(Carnot)}}$ becomes large when $T_h\approx T_c$, diverging when $T_h=T_c$. This explains why we observe a refrigerator at $T_f\gtrsim T_i$ in Fig.~\ref{sec:results;sub:operations;fig:diagram}. In contrast, the efficiency $\eta^{\text{(Carnot)}}$ increases instead with a large temperature difference between the reservoirs.   
 
\subsubsection{Dependence on the temperatures}

Firstly, we examine the refrigerator's performance as a function of the temperatures for strong interactions. We show such performance in Fig.~\ref{sec:results;sub:refrigerator;fig:T}, corresponding to the same region reported in Fig.~\ref{sec:results;sub:operations;fig:diagram}. For fixed interactions [panel (a)], we observe that the refrigerator operates at an approximately constant performance of $\COPR\approx 0.2$ in the whole plotted region. This seems to mirror the engine's behavior for this protocol, which also shows a fairly constant efficiency [panel (a) of Fig.~\ref{sec:results;sub:engine;fig:T}].

A stronger dependence on the temperatures is displayed by the increasing interaction protocol [panel (b)]. Here, we observe a larger performance of approximately $\COPR\approx 1$ at small temperature differences $T_f\approx T_i$, corresponding to the light yellow region. However, such a small temperature difference between reservoirs might defeat the purpose of a refrigerator machine. Nevertheless, while $\COPR$ decreases for increasing $T_f/T_i$, the performance of the increasing protocol is generally larger than that for fixed interactions.

Finally, for decreasing interactions [panel (c)], the figure displays only a very small region of operation with a near-vanishing $\COPR$. Thus, these results support our previous conclusions that the increasing interaction protocol favors refrigeration operations, while the decreasing protocol hinders it.

\subsubsection{Dependence on the interaction}

\begin{figure*}[t!]
\centering
\includegraphics[width=\textwidth]{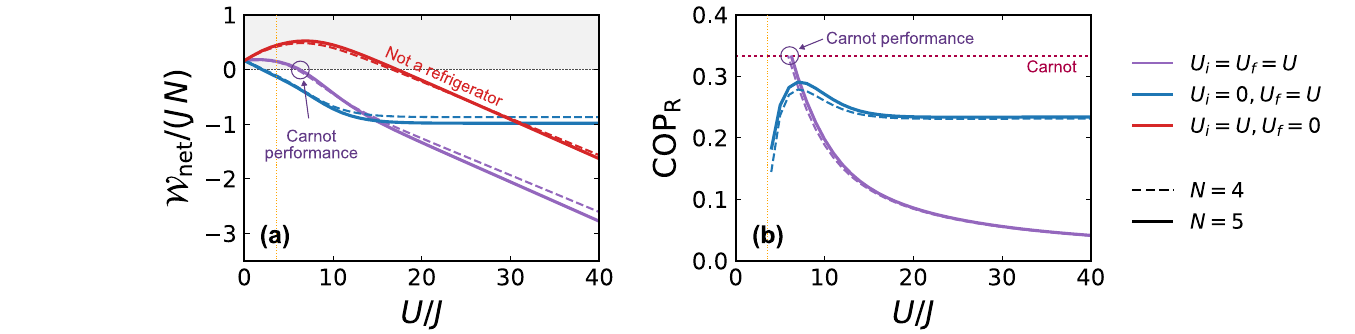}
\caption{Net work output per particle (a) and coefficient of performance of the refrigeration cycle (b) as a function of the protocol's interaction $U$. 
The calculations consider an initial lattice size of $M_i/N=2$ and a final lattice size of $M_f/N=1$. The engine operates between reservoirs with temperatures $k_B T_i/J=0.5$ and $k_B T_f/J=2.0$. Each curve considers the protocol of interactions and number of particles indicated in the legend. The horizontal dotted line in (b) indicates the Carnot coefficient of performance. The vertical orange dotted lines in both panels indicate the SF–MI phase transition interaction in the thermodynamic limit $U/J\approx 3.6$~\cite{kuhner_phases_1998}.}
\label{sec:results;sub:refrigerator;fig:U}
\end{figure*}

Finally, we examine the performance of the refrigeration cycle as a function of the interaction strength. We show a representative example in Fig.~\ref{sec:results;sub:refrigerator;fig:U}. Here, for simplicity, we continue focusing on compression to unit filling. Additionally, we focus on a temperature difference with a factor of four between the two reservoirs, instead of $T_f\approx T_i$.

The work output [panel (a)] becomes negative at large $U$ for the three protocols. However, the cycle operates as a refrigerator only for the fixed (purple curves) and increasing (blue curves) protocols. In contrast, the decreasing interaction protocol (red curves) does not operate as a refrigerator despite showing $\Wnet<0$ for large repulsion. We also observe that while the fixed and increasing interaction protocols have a similar work input for $U/J\lesssim 15$, for fixed interactions $\Wnet$ continues to decrease linearly for large $U$ (see purple curves in the bottom-right region of the panel). This means that the energy cost becomes very large for extremely strong repulsion. In contrast, for increasing interactions we observe that $\Wnet$ saturates to a constant value of $\Wnet/(JN)\approx -1$ for large $U$, showing that this protocol displays a manageable energy cost.

The previous results are supported by the coefficient of performance [panel (b)]. For the considered temperatures, the values of $\COPR$ for both working protocols are comparable around $U/J\approx 10$. However, $\COPR$ decreases strongly as $U$ increases for fixed interactions. In contrast, for increasing interactions we observe that the coefficient of performance saturates to $\COPR\approx 0.25$ for $U\to\infty$, further confirming that this protocol favors higher-performance refrigeration cycles.

Finally, we note that for fixed interactions, the coefficient of performance approaches the Carnot limit when $\Wnet=0$, as indicated by the purple circles. This is the expected result for an Otto cycle.

\section{Conclusions}
\label{sec:concl}

In this work, we proposed a controllable quantum thermal machine that utilizes ultracold interacting bosons in an optical lattice as the working medium. Our proposal features an Otto-inspired cycle with strokes that alter the interatomic interactions, in addition to strokes that change the number of available sites in the lattice to simulate compression and expansion processes. Our results show that interactions and lattice sizes can be manipulated as thermodynamic resources, leading to enhanced performance over non-interacting Otto cycles.

One of the main results of this work is that compressing the medium to a Mott insulator state can greatly improve engine performance. In particular, the highest performance, meaning both a large work output and high efficiency, is achieved by reducing the interaction after the compression and heating strokes. Furthermore, such a protocol broadens the range of temperatures at which the engine operation is possible. We found that this performance enhancement is possible because additional heat can be absorbed when the medium is an insulator.

Additionally, our thermal machine can display other operating regimes depending on the interaction and temperature changes. In particular, increasing the interactions after the compression strokes generally favors refrigeration cycles, being able to achieve near-Carnot performance. Overall, these results reveal a marked asymmetry between the two directions of interaction changes. Decreasing interactions after compression favors heat-engine operations, whereas increasing interactions favors refrigeration. This asymmetry serves as an additional control knob for selecting the thermodynamic function of the same working medium.

In summary, our findings further show that interactions and strongly correlated lattice regimes can be harnessed as useful thermodynamic resources. While this work considered a minimal medium with a few bosons, the proposed setup could be realized in conventional experiments of ultracold atoms in optical lattices. On the theoretical side, a natural extension of this work includes the consideration of harmonic confinement, which would enable us to perform the compression/expansion strokes through changes in the trap's frequency instead of in the available sites.  Future work could also consider a larger working medium and higher-dimensional geometries, which could be studied numerically using Quantum Monte Carlo simulations~\cite{wessel_quantum_2004} or finite-temperature matrix product operators~\cite{verstraete_matrix_2004,feiguin_finite-temperature_2005}. Additionally, it would be interesting to investigate finite-time implementations of the lattice-compression and interaction strokes, as non-adiabatic excitations could further alter the machine's practical performance.

\section{Acknowledgments}
F. I. and P. V. acknowledge support from CEDENNA under Grant CIA No. 250002.  F.J.P. acknowledges support from FONDECYT (Chile) under Grant No. 1250173 and financial support from the Brazilian National Council for Scientific and Technological Development (CNPq), under Grant No. 446761/2025-7. P.V. also acknowledges support from FONDECYT (Chile) under Grant No. 1240582. T.F. acknowledges support from JSPS KAKENHI Grant No. JP23K03290 and JST Grant No. JPMJPF2221.

\appendix

\section{Properties of the working medium}
\label{app:thermo}

\begin{figure*}[t!]
\centering
\includegraphics[width=\textwidth]{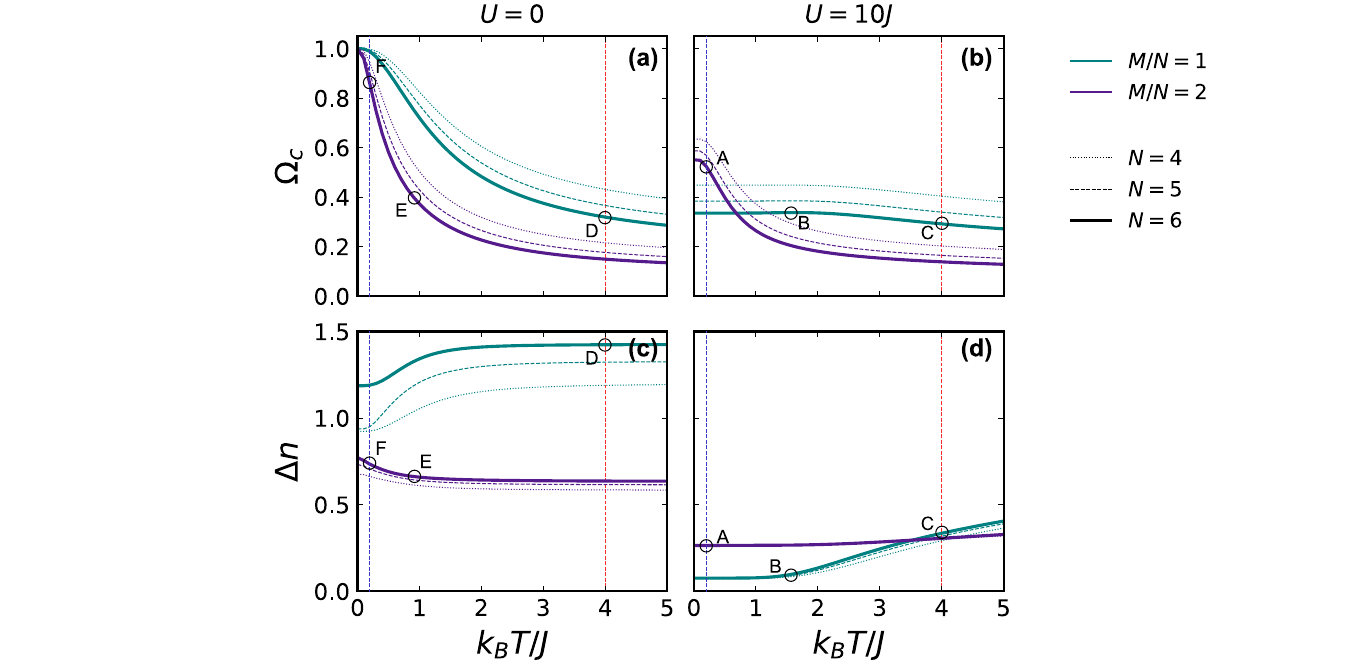}
\caption{Condensate fraction [(a) and (b)] and density variance [(c) and (d)] as a function of temperature. Each curve shows the lattice size and the number of particles indicated in the legend, while each column of panels shows the interaction strengths indicated above the figure. The vertical dashed blue lines on the left of each panel indicate $k_B T/J=0.2$, while the vertical dashed red lines on the right indicate $k_BT/J =4.0$. The circles indicate the states in a cycle with the decreasing interaction protocol for $k_B T_i/J=0.2$ and $k_B T_f/J=4.0$, corresponding to the cycle highlighted with the orange circles in Fig.~\ref{sec:results;sub:engine;fig:M}(c) and (f).}
\label{app:thermo;fig:bosehubbard}
\end{figure*}

In this appendix, we examine whether the bosons are in a superfluid, insulating, or normal state. Even though our small finite system shows crossovers instead of well-defined phase transitions, we can still qualitatively estimate the overall state of the system. To do so, we calculate the condensate fraction $\Omega_c=\lambda_{\text{max}}/N$, where $\lambda_{\text{max}}$ is the largest eigenvalue of the single-particle density matrix $\rho^{\text{(sp)}}$~\cite{zhang_exact_2010}. The latter is defined as
\begin{equation}
    \rho_{i,j}^{\text{(sp)}}=\langle \aopd_{i}\hat{a}_j\rangle.
\end{equation}
The condensate fraction takes values $0\leq \Omega_c\leq 1$. We can identify an SF state when the condensate fraction is close to one, $\Omega_c\approx 1$. In contrast, $\Omega_c$ is small in the MI and normal states. To differentiate the remaining two states, we calculate the density variance~\cite{zhang_exact_2010}
\begin{equation}
\Delta n=\frac{1}{M}\sum_{i=1}^M\left(\langle \nop_i^2\rangle-\langle \nop_i\rangle^2\right).
\end{equation}
This is small, $\Delta n\approx 0$, in the MI state at unit filling, as each site is occupied by a single boson [see the bottom illustration in Fig.~\ref{sec:model;fig:illustration}(a)]. In contrast, $\Delta n$ becomes larger in other states, as sites can be occupied by several bosons.

We show a representative set of these quantities as a function of temperature in Fig.~\ref{app:thermo;fig:bosehubbard}. Additionally, we mark with circles the states of reference of a representative cycle with the decreasing interaction protocol from the main text. We observe that $\Omega_c\approx 1$ at low temperatures in the non-interacting limit [upper-left region in panel (a)], confirming that they correspond to SF bosons regardless of the filling factor (similar results are obtained for small $U$). In particular, we observe that the state of reference F is deep in the SF regime. Then, as $T$ increases, $\Omega_c$ decreases rapidly, taking values $\Omega_c\lesssim 0.2$ for $k_B T/J>2$. This shows that at such high temperatures the superfluid state is destroyed, as expected. In turn, $\Omega_c$ is small at low temperatures for strong repulsion and at unit filling [teal curves in (b)], as it corresponds to an MI state. The latter is further confirmed by the corresponding small value of $\Delta n$ at low $T$ [teal curves in panel (d)]. Importantly, we observe that state B clearly lies within the MI regime, confirming that we performed compressions to the MI state in the main text. In contrast, $\Delta n$ is larger for no interactions [panel (c)] or at high temperatures. Thus, the MI state is destroyed at $k_B T/J\gg 1$, and thus C and D lie deep within the normal regime.

\section{General dependence on the interactions}
\label{app:interactions}

\begin{figure*}[t!]
\centering
\includegraphics[width=\textwidth]{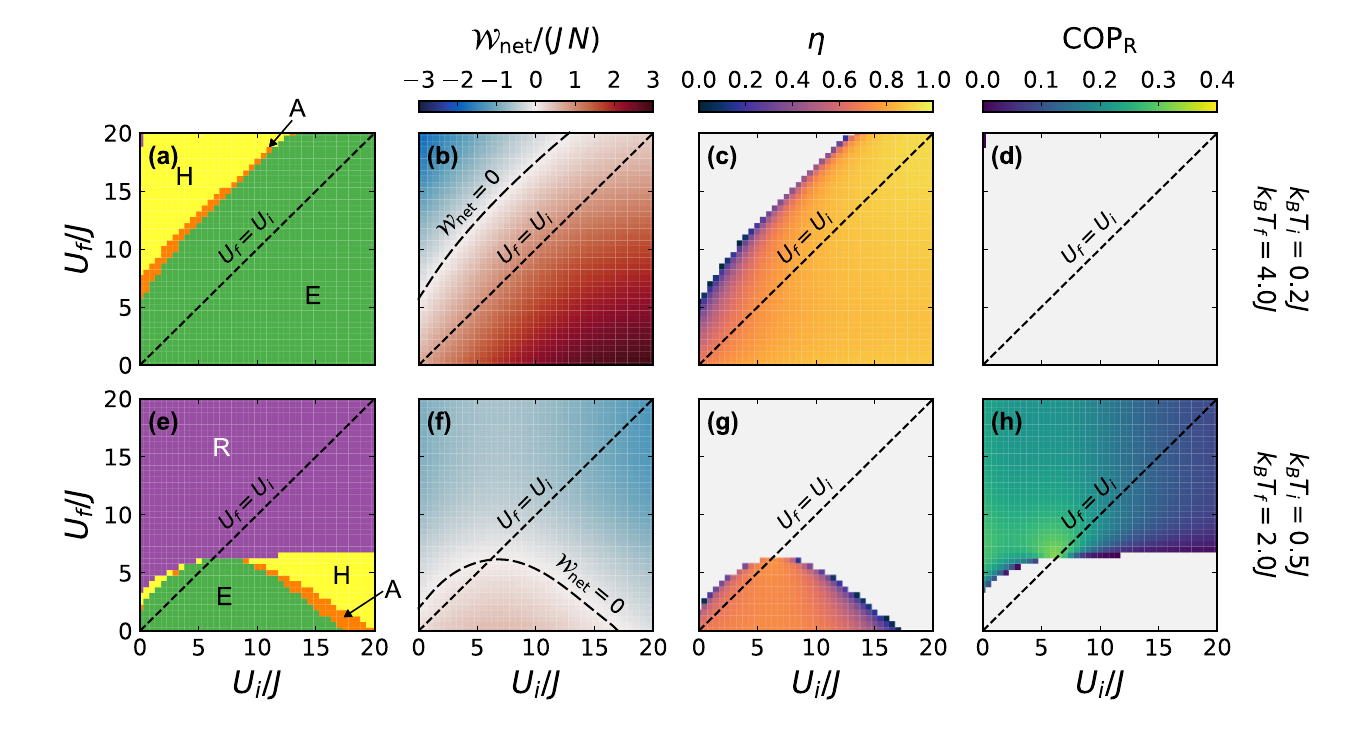}
\caption{Thermal operations [(a) and (e)], net work output per particle [(b) and (f)], engine efficiency [(c) and (g)], and coefficient of performance [(d) and (h)] as functions of the initial and final interaction strengths $U_i$ and $U_f$. 
The calculations consider $N=5$, an initial lattice size of $M_i/N=2$, and a final lattice size of $M_f/N=1$. Each row of panels considers the temperatures indicated on the right of the figure. The gray regions indicate that the thermal machine does not operate as an engine [panels (c) and (g)] or as a refrigerator [panels (d) and (h)].}
\label{app:interactions;fig:performance}
\end{figure*}

Throughout the main text, we examined cycles with either fixed interactions or changes between $U=0$ and a finite $U$. To examine how the cycle behaves for arbitrary interaction strengths, in Fig.~\ref{app:interactions;fig:performance} we show the operational modes (left panels) as a function of $U_i$ and $U_f$ for compression to unit filling. Additionally, the figure displays the corresponding work output and performance of both the engine and refrigeration cycles. In the figure, the top panels consider a large temperature difference between reservoirs, favoring engine operation. In contrast, the bottom panels consider a smaller temperature difference, favoring the refrigeration cycle. We note that the diagonals in the panels correspond to the protocol of equal interactions $U_i=U_f$.  In turn, the left $y$-axes correspond to the limit of the increasing interaction protocol of the main text ($U_i=0$ and $U_f>0$), while the bottom $x$-axes correspond to the limit of the decreasing interaction protocol ($U_i>0$ and $U_f=0$)

For large temperature differences [panel (a)], we observe that the machine operates as an engine for a large range of interactions in the bottom-right region of the panel, corresponding mostly to decreasing interactions. This further supports the conclusions of the main text. In contrast, for these temperatures, the cycle behaves as a heater for large decreasing interactions, with a narrow accelerator band between the heater and engine operations. Conversely, the cycle reaches a large work output [panel (b)] and large efficiency [panel (c)] within the engine's operational regime. Naturally, there is no $\COPR$ [panel (d)], as the cycle does not behave as a refrigerator for these temperatures.

For smaller temperature differences [panel (e)], the cycle displays the four operations. The machine operates as a refrigerator for a wide range of interactions, particularly for increasing interactions (top left region), also supporting the conclusions of the main text. Importantly, while the work output  [panel (f)] remains approximately constant within the refrigeration region, the coefficient of performance [panel (h)] is slightly larger in the increasing-interaction half of the panel. In turn, the other three operations appear approximately at $U_f/J\lesssim 5$. For the engine, this is consistent with the behavior shown in Fig.~\ref{sec:results;sub:engine;fig:U}(a).

\newpage
\bibliography{biblio}

\begin{thebibliography}{53}%
\makeatletter
\providecommand \@ifxundefined [1]{%
 \@ifx{#1\undefined}
}%
\providecommand \@ifnum [1]{%
 \ifnum #1\expandafter \@firstoftwo
 \else \expandafter \@secondoftwo
 \fi
}%
\providecommand \@ifx [1]{%
 \ifx #1\expandafter \@firstoftwo
 \else \expandafter \@secondoftwo
 \fi
}%
\providecommand \natexlab [1]{#1}%
\providecommand \enquote  [1]{``#1''}%
\providecommand \bibnamefont  [1]{#1}%
\providecommand \bibfnamefont [1]{#1}%
\providecommand \citenamefont [1]{#1}%
\providecommand \href@noop [0]{\@secondoftwo}%
\providecommand \href [0]{\begingroup \@sanitize@url \@href}%
\providecommand \@href[1]{\@@startlink{#1}\@@href}%
\providecommand \@@href[1]{\endgroup#1\@@endlink}%
\providecommand \@sanitize@url [0]{\catcode `\\12\catcode `\$12\catcode `\&12\catcode `\#12\catcode `\^12\catcode `\_12\catcode `\%12\relax}%
\providecommand \@@startlink[1]{}%
\providecommand \@@endlink[0]{}%
\providecommand \url  [0]{\begingroup\@sanitize@url \@url }%
\providecommand \@url [1]{\endgroup\@href {#1}{\urlprefix }}%
\providecommand \urlprefix  [0]{URL }%
\providecommand \Eprint [0]{\href }%
\providecommand \doibase [0]{https://doi.org/}%
\providecommand \selectlanguage [0]{\@gobble}%
\providecommand \bibinfo  [0]{\@secondoftwo}%
\providecommand \bibfield  [0]{\@secondoftwo}%
\providecommand \translation [1]{[#1]}%
\providecommand \BibitemOpen [0]{}%
\providecommand \bibitemStop [0]{}%
\providecommand \bibitemNoStop [0]{.\EOS\space}%
\providecommand \EOS [0]{\spacefactor3000\relax}%
\providecommand \BibitemShut  [1]{\csname bibitem#1\endcsname}%
\let\auto@bib@innerbib\@empty
\bibitem [{\citenamefont {Kosloff}\ and\ \citenamefont {Levy}(2014)}]{kosloff_quantum_2014}%
  \BibitemOpen
  \bibfield  {author} {\bibinfo {author} {\bibfnamefont {R.}~\bibnamefont {Kosloff}}\ and\ \bibinfo {author} {\bibfnamefont {A.}~\bibnamefont {Levy}},\ }\bibfield  {title} {\bibinfo {title} {Quantum {Heat} {Engines} and {Refrigerators}: {Continuous} {Devices}},\ }\href {https://doi.org/10.1146/annurev-physchem-040513-103724} {\bibfield  {journal} {\bibinfo  {journal} {Annual Review of Physical Chemistry}\ }\textbf {\bibinfo {volume} {65}},\ \bibinfo {pages} {365} (\bibinfo {year} {2014})}\BibitemShut {NoStop}%
\bibitem [{\citenamefont {Vinjanampathy}\ and\ \citenamefont {Anders}(2016)}]{vinjanampathy_quantum_2016}%
  \BibitemOpen
  \bibfield  {author} {\bibinfo {author} {\bibfnamefont {S.}~\bibnamefont {Vinjanampathy}}\ and\ \bibinfo {author} {\bibfnamefont {J.}~\bibnamefont {Anders}},\ }\bibfield  {title} {\bibinfo {title} {Quantum thermodynamics},\ }\href {https://doi.org/10.1080/00107514.2016.1201896} {\bibfield  {journal} {\bibinfo  {journal} {Contemporary Physics}\ }\textbf {\bibinfo {volume} {57}},\ \bibinfo {pages} {545} (\bibinfo {year} {2016})}\BibitemShut {NoStop}%
\bibitem [{\citenamefont {Myers}\ \emph {et~al.}(2022{\natexlab{a}})\citenamefont {Myers}, \citenamefont {Abah},\ and\ \citenamefont {Deffner}}]{myers_quantum_2022}%
  \BibitemOpen
  \bibfield  {author} {\bibinfo {author} {\bibfnamefont {N.~M.}\ \bibnamefont {Myers}}, \bibinfo {author} {\bibfnamefont {O.}~\bibnamefont {Abah}},\ and\ \bibinfo {author} {\bibfnamefont {S.}~\bibnamefont {Deffner}},\ }\bibfield  {title} {\bibinfo {title} {Quantum thermodynamic devices: {From} theoretical proposals to experimental reality},\ }\href {https://doi.org/10.1116/5.0083192} {\bibfield  {journal} {\bibinfo  {journal} {AVS Quantum Science}\ }\textbf {\bibinfo {volume} {4}},\ \bibinfo {pages} {027101} (\bibinfo {year} {2022}{\natexlab{a}})}\BibitemShut {NoStop}%
\bibitem [{\citenamefont {Cangemi}\ \emph {et~al.}(2024)\citenamefont {Cangemi}, \citenamefont {Bhadra},\ and\ \citenamefont {Levy}}]{cangemi_quantum_2024}%
  \BibitemOpen
  \bibfield  {author} {\bibinfo {author} {\bibfnamefont {L.~M.}\ \bibnamefont {Cangemi}}, \bibinfo {author} {\bibfnamefont {C.}~\bibnamefont {Bhadra}},\ and\ \bibinfo {author} {\bibfnamefont {A.}~\bibnamefont {Levy}},\ }\bibfield  {title} {\bibinfo {title} {Quantum engines and refrigerators},\ }\href {https://doi.org/10.1016/j.physrep.2024.07.001} {\bibfield  {journal} {\bibinfo  {journal} {Physics Reports}\ }\textbf {\bibinfo {volume} {1087}},\ \bibinfo {pages} {1} (\bibinfo {year} {2024})}\BibitemShut {NoStop}%
\bibitem [{\citenamefont {Potts}(2026)}]{potts_quantum_2026}%
  \BibitemOpen
  \bibfield  {author} {\bibinfo {author} {\bibfnamefont {P.}~\bibnamefont {Potts}},\ }\bibfield  {title} {\bibinfo {title} {Quantum thermodynamics},\ }\href {https://doi.org/10.21468/SciPostPhysLectNotes.126} {\bibfield  {journal} {\bibinfo  {journal} {SciPost Physics Lecture Notes}\ ,\ \bibinfo {pages} {126}} (\bibinfo {year} {2026})}\BibitemShut {NoStop}%
\bibitem [{\citenamefont {Campisi}\ and\ \citenamefont {Fazio}(2016)}]{campisi_power_2016}%
  \BibitemOpen
  \bibfield  {author} {\bibinfo {author} {\bibfnamefont {M.}~\bibnamefont {Campisi}}\ and\ \bibinfo {author} {\bibfnamefont {R.}~\bibnamefont {Fazio}},\ }\bibfield  {title} {\bibinfo {title} {The power of a critical heat engine},\ }\href {https://doi.org/10.1038/ncomms11895} {\bibfield  {journal} {\bibinfo  {journal} {Nature Communications}\ }\textbf {\bibinfo {volume} {7}},\ \bibinfo {pages} {11895} (\bibinfo {year} {2016})}\BibitemShut {NoStop}%
\bibitem [{\citenamefont {Ma}\ \emph {et~al.}(2017)\citenamefont {Ma}, \citenamefont {Su},\ and\ \citenamefont {Sun}}]{ma_quantum_2017}%
  \BibitemOpen
  \bibfield  {author} {\bibinfo {author} {\bibfnamefont {Y.-H.}\ \bibnamefont {Ma}}, \bibinfo {author} {\bibfnamefont {S.-H.}\ \bibnamefont {Su}},\ and\ \bibinfo {author} {\bibfnamefont {C.-P.}\ \bibnamefont {Sun}},\ }\bibfield  {title} {\bibinfo {title} {Quantum thermodynamic cycle with quantum phase transition},\ }\href {https://doi.org/10.1103/PhysRevE.96.022143} {\bibfield  {journal} {\bibinfo  {journal} {Physical Review E}\ }\textbf {\bibinfo {volume} {96}},\ \bibinfo {pages} {022143} (\bibinfo {year} {2017})}\BibitemShut {NoStop}%
\bibitem [{\citenamefont {Chand}\ and\ \citenamefont {Biswas}(2018)}]{chand_critical-point_2018}%
  \BibitemOpen
  \bibfield  {author} {\bibinfo {author} {\bibfnamefont {S.}~\bibnamefont {Chand}}\ and\ \bibinfo {author} {\bibfnamefont {A.}~\bibnamefont {Biswas}},\ }\bibfield  {title} {\bibinfo {title} {Critical-point behavior of a measurement-based quantum heat engine},\ }\href {https://doi.org/10.1103/PhysRevE.98.052147} {\bibfield  {journal} {\bibinfo  {journal} {Physical Review E}\ }\textbf {\bibinfo {volume} {98}},\ \bibinfo {pages} {052147} (\bibinfo {year} {2018})}\BibitemShut {NoStop}%
\bibitem [{\citenamefont {B.~S}\ \emph {et~al.}(2020)\citenamefont {B.~S}, \citenamefont {Mukherjee}, \citenamefont {Divakaran},\ and\ \citenamefont {Del~Campo}}]{b_s_universal_2020}%
  \BibitemOpen
  \bibfield  {author} {\bibinfo {author} {\bibfnamefont {R.}~\bibnamefont {B.~S}}, \bibinfo {author} {\bibfnamefont {V.}~\bibnamefont {Mukherjee}}, \bibinfo {author} {\bibfnamefont {U.}~\bibnamefont {Divakaran}},\ and\ \bibinfo {author} {\bibfnamefont {A.}~\bibnamefont {Del~Campo}},\ }\bibfield  {title} {\bibinfo {title} {Universal finite-time thermodynamics of many-body quantum machines from {Kibble}-{Zurek} scaling},\ }\href {https://doi.org/10.1103/PhysRevResearch.2.043247} {\bibfield  {journal} {\bibinfo  {journal} {Physical Review Research}\ }\textbf {\bibinfo {volume} {2}},\ \bibinfo {pages} {043247} (\bibinfo {year} {2020})}\BibitemShut {NoStop}%
\bibitem [{\citenamefont {Myers}\ \emph {et~al.}(2022{\natexlab{b}})\citenamefont {Myers}, \citenamefont {Peña}, \citenamefont {Negrete}, \citenamefont {Vargas}, \citenamefont {De~Chiara},\ and\ \citenamefont {Deffner}}]{myers_boosting_2022}%
  \BibitemOpen
  \bibfield  {author} {\bibinfo {author} {\bibfnamefont {N.~M.}\ \bibnamefont {Myers}}, \bibinfo {author} {\bibfnamefont {F.~J.}\ \bibnamefont {Peña}}, \bibinfo {author} {\bibfnamefont {O.}~\bibnamefont {Negrete}}, \bibinfo {author} {\bibfnamefont {P.}~\bibnamefont {Vargas}}, \bibinfo {author} {\bibfnamefont {G.}~\bibnamefont {De~Chiara}},\ and\ \bibinfo {author} {\bibfnamefont {S.}~\bibnamefont {Deffner}},\ }\bibfield  {title} {\bibinfo {title} {Boosting engine performance with {Bose}–{Einstein} condensation},\ }\href {https://doi.org/10.1088/1367-2630/ac47cc} {\bibfield  {journal} {\bibinfo  {journal} {New Journal of Physics}\ }\textbf {\bibinfo {volume} {24}},\ \bibinfo {pages} {025001} (\bibinfo {year} {2022}{\natexlab{b}})}\BibitemShut {NoStop}%
\bibitem [{\citenamefont {Zheng}\ and\ \citenamefont {Poletti}(2015)}]{zheng_quantum_2015}%
  \BibitemOpen
  \bibfield  {author} {\bibinfo {author} {\bibfnamefont {Y.}~\bibnamefont {Zheng}}\ and\ \bibinfo {author} {\bibfnamefont {D.}~\bibnamefont {Poletti}},\ }\bibfield  {title} {\bibinfo {title} {Quantum statistics and the performance of engine cycles},\ }\href {https://doi.org/10.1103/PhysRevE.92.012110} {\bibfield  {journal} {\bibinfo  {journal} {Physical Review E}\ }\textbf {\bibinfo {volume} {92}},\ \bibinfo {pages} {012110} (\bibinfo {year} {2015})}\BibitemShut {NoStop}%
\bibitem [{\citenamefont {Myers}\ and\ \citenamefont {Deffner}(2020)}]{myers_bosons_2020}%
  \BibitemOpen
  \bibfield  {author} {\bibinfo {author} {\bibfnamefont {N.~M.}\ \bibnamefont {Myers}}\ and\ \bibinfo {author} {\bibfnamefont {S.}~\bibnamefont {Deffner}},\ }\bibfield  {title} {\bibinfo {title} {Bosons outperform fermions: {The} thermodynamic advantage of symmetry},\ }\href {https://doi.org/10.1103/PhysRevE.101.012110} {\bibfield  {journal} {\bibinfo  {journal} {Physical Review E}\ }\textbf {\bibinfo {volume} {101}},\ \bibinfo {pages} {012110} (\bibinfo {year} {2020})}\BibitemShut {NoStop}%
\bibitem [{\citenamefont {Watanabe}\ \emph {et~al.}(2020)\citenamefont {Watanabe}, \citenamefont {Venkatesh}, \citenamefont {Talkner}, \citenamefont {Hwang},\ and\ \citenamefont {Del~Campo}}]{watanabe_quantum_2020}%
  \BibitemOpen
  \bibfield  {author} {\bibinfo {author} {\bibfnamefont {G.}~\bibnamefont {Watanabe}}, \bibinfo {author} {\bibfnamefont {B.~P.}\ \bibnamefont {Venkatesh}}, \bibinfo {author} {\bibfnamefont {P.}~\bibnamefont {Talkner}}, \bibinfo {author} {\bibfnamefont {M.-J.}\ \bibnamefont {Hwang}},\ and\ \bibinfo {author} {\bibfnamefont {A.}~\bibnamefont {Del~Campo}},\ }\bibfield  {title} {\bibinfo {title} {Quantum {Statistical} {Enhancement} of the {Collective} {Performance} of {Multiple} {Bosonic} {Engines}},\ }\href {https://doi.org/10.1103/PhysRevLett.124.210603} {\bibfield  {journal} {\bibinfo  {journal} {Physical Review Letters}\ }\textbf {\bibinfo {volume} {124}},\ \bibinfo {pages} {210603} (\bibinfo {year} {2020})}\BibitemShut {NoStop}%
\bibitem [{\citenamefont {Myers}\ and\ \citenamefont {Deffner}(2021)}]{myers_thermodynamics_2021}%
  \BibitemOpen
  \bibfield  {author} {\bibinfo {author} {\bibfnamefont {N.~M.}\ \bibnamefont {Myers}}\ and\ \bibinfo {author} {\bibfnamefont {S.}~\bibnamefont {Deffner}},\ }\bibfield  {title} {\bibinfo {title} {Thermodynamics of {Statistical} {Anyons}},\ }\href {https://doi.org/10.1103/PRXQuantum.2.040312} {\bibfield  {journal} {\bibinfo  {journal} {PRX Quantum}\ }\textbf {\bibinfo {volume} {2}},\ \bibinfo {pages} {040312} (\bibinfo {year} {2021})}\BibitemShut {NoStop}%
\bibitem [{\citenamefont {Dunlop}\ \emph {et~al.}(2026)\citenamefont {Dunlop}, \citenamefont {Tejero}, \citenamefont {Skotiniotis},\ and\ \citenamefont {Manzano}}]{dunlop_thermodynamics_2026}%
  \BibitemOpen
  \bibfield  {author} {\bibinfo {author} {\bibfnamefont {J.}~\bibnamefont {Dunlop}}, \bibinfo {author} {\bibfnamefont {A.}~\bibnamefont {Tejero}}, \bibinfo {author} {\bibfnamefont {M.}~\bibnamefont {Skotiniotis}},\ and\ \bibinfo {author} {\bibfnamefont {D.}~\bibnamefont {Manzano}},\ }\href {https://doi.org/10.48550/arXiv.2502.19019} {\bibinfo {title} {Thermodynamics of {Hamiltonian} anyons with applications to quantum heat engines}} (\bibinfo {year} {2026}),\ \bibinfo {note} {arXiv:2502.19019}\BibitemShut {NoStop}%
\bibitem [{\citenamefont {Huang}\ \emph {et~al.}(2012)\citenamefont {Huang}, \citenamefont {Wang},\ and\ \citenamefont {Yi}}]{huang_effects_2012}%
  \BibitemOpen
  \bibfield  {author} {\bibinfo {author} {\bibfnamefont {X.~L.}\ \bibnamefont {Huang}}, \bibinfo {author} {\bibfnamefont {T.}~\bibnamefont {Wang}},\ and\ \bibinfo {author} {\bibfnamefont {X.~X.}\ \bibnamefont {Yi}},\ }\bibfield  {title} {\bibinfo {title} {Effects of reservoir squeezing on quantum systems and work extraction},\ }\href {https://doi.org/10.1103/PhysRevE.86.051105} {\bibfield  {journal} {\bibinfo  {journal} {Physical Review E}\ }\textbf {\bibinfo {volume} {86}},\ \bibinfo {pages} {051105} (\bibinfo {year} {2012})}\BibitemShut {NoStop}%
\bibitem [{\citenamefont {Roßnagel}\ \emph {et~al.}(2014)\citenamefont {Roßnagel}, \citenamefont {Abah}, \citenamefont {Schmidt-Kaler}, \citenamefont {Singer},\ and\ \citenamefont {Lutz}}]{rosnagel_nanoscale_2014}%
  \BibitemOpen
  \bibfield  {author} {\bibinfo {author} {\bibfnamefont {J.}~\bibnamefont {Roßnagel}}, \bibinfo {author} {\bibfnamefont {O.}~\bibnamefont {Abah}}, \bibinfo {author} {\bibfnamefont {F.}~\bibnamefont {Schmidt-Kaler}}, \bibinfo {author} {\bibfnamefont {K.}~\bibnamefont {Singer}},\ and\ \bibinfo {author} {\bibfnamefont {E.}~\bibnamefont {Lutz}},\ }\bibfield  {title} {\bibinfo {title} {Nanoscale {Heat} {Engine} {Beyond} the {Carnot} {Limit}},\ }\href {https://doi.org/10.1103/PhysRevLett.112.030602} {\bibfield  {journal} {\bibinfo  {journal} {Physical Review Letters}\ }\textbf {\bibinfo {volume} {112}},\ \bibinfo {pages} {030602} (\bibinfo {year} {2014})}\BibitemShut {NoStop}%
\bibitem [{\citenamefont {Manzano}\ \emph {et~al.}(2016)\citenamefont {Manzano}, \citenamefont {Galve}, \citenamefont {Zambrini},\ and\ \citenamefont {Parrondo}}]{manzano_entropy_2016}%
  \BibitemOpen
  \bibfield  {author} {\bibinfo {author} {\bibfnamefont {G.}~\bibnamefont {Manzano}}, \bibinfo {author} {\bibfnamefont {F.}~\bibnamefont {Galve}}, \bibinfo {author} {\bibfnamefont {R.}~\bibnamefont {Zambrini}},\ and\ \bibinfo {author} {\bibfnamefont {J.~M.~R.}\ \bibnamefont {Parrondo}},\ }\bibfield  {title} {\bibinfo {title} {Entropy production and thermodynamic power of the squeezed thermal reservoir},\ }\href {https://doi.org/10.1103/PhysRevE.93.052120} {\bibfield  {journal} {\bibinfo  {journal} {Physical Review E}\ }\textbf {\bibinfo {volume} {93}},\ \bibinfo {pages} {052120} (\bibinfo {year} {2016})}\BibitemShut {NoStop}%
\bibitem [{\citenamefont {Klaers}\ \emph {et~al.}(2017)\citenamefont {Klaers}, \citenamefont {Faelt}, \citenamefont {Imamoglu},\ and\ \citenamefont {Togan}}]{klaers_squeezed_2017}%
  \BibitemOpen
  \bibfield  {author} {\bibinfo {author} {\bibfnamefont {J.}~\bibnamefont {Klaers}}, \bibinfo {author} {\bibfnamefont {S.}~\bibnamefont {Faelt}}, \bibinfo {author} {\bibfnamefont {A.}~\bibnamefont {Imamoglu}},\ and\ \bibinfo {author} {\bibfnamefont {E.}~\bibnamefont {Togan}},\ }\bibfield  {title} {\bibinfo {title} {Squeezed {Thermal} {Reservoirs} as a {Resource} for a {Nanomechanical} {Engine} beyond the {Carnot} {Limit}},\ }\href {https://doi.org/10.1103/PhysRevX.7.031044} {\bibfield  {journal} {\bibinfo  {journal} {Physical Review X}\ }\textbf {\bibinfo {volume} {7}},\ \bibinfo {pages} {031044} (\bibinfo {year} {2017})}\BibitemShut {NoStop}%
\bibitem [{\citenamefont {Jaramillo}\ \emph {et~al.}(2016)\citenamefont {Jaramillo}, \citenamefont {Beau},\ and\ \citenamefont {Campo}}]{jaramillo_quantum_2016}%
  \BibitemOpen
  \bibfield  {author} {\bibinfo {author} {\bibfnamefont {J.}~\bibnamefont {Jaramillo}}, \bibinfo {author} {\bibfnamefont {M.}~\bibnamefont {Beau}},\ and\ \bibinfo {author} {\bibfnamefont {A.~D.}\ \bibnamefont {Campo}},\ }\bibfield  {title} {\bibinfo {title} {Quantum supremacy of many-particle thermal machines},\ }\href {https://doi.org/10.1088/1367-2630/18/7/075019} {\bibfield  {journal} {\bibinfo  {journal} {New Journal of Physics}\ }\textbf {\bibinfo {volume} {18}},\ \bibinfo {pages} {075019} (\bibinfo {year} {2016})}\BibitemShut {NoStop}%
\bibitem [{\citenamefont {Niedenzu}\ and\ \citenamefont {Kurizki}(2018)}]{niedenzu_cooperative_2018}%
  \BibitemOpen
  \bibfield  {author} {\bibinfo {author} {\bibfnamefont {W.}~\bibnamefont {Niedenzu}}\ and\ \bibinfo {author} {\bibfnamefont {G.}~\bibnamefont {Kurizki}},\ }\bibfield  {title} {\bibinfo {title} {Cooperative many-body enhancement of quantum thermal machine power},\ }\href {https://doi.org/10.1088/1367-2630/aaed55} {\bibfield  {journal} {\bibinfo  {journal} {New Journal of Physics}\ }\textbf {\bibinfo {volume} {20}},\ \bibinfo {pages} {113038} (\bibinfo {year} {2018})}\BibitemShut {NoStop}%
\bibitem [{\citenamefont {Chen}\ \emph {et~al.}(2018)\citenamefont {Chen}, \citenamefont {Dong},\ and\ \citenamefont {Sun}}]{chen_bose-fermi_2018}%
  \BibitemOpen
  \bibfield  {author} {\bibinfo {author} {\bibfnamefont {J.}~\bibnamefont {Chen}}, \bibinfo {author} {\bibfnamefont {H.}~\bibnamefont {Dong}},\ and\ \bibinfo {author} {\bibfnamefont {C.-P.}\ \bibnamefont {Sun}},\ }\bibfield  {title} {\bibinfo {title} {Bose-{Fermi} duality in a quantum {Otto} heat engine with trapped repulsive bosons},\ }\href {https://doi.org/10.1103/PhysRevE.98.062119} {\bibfield  {journal} {\bibinfo  {journal} {Physical Review E}\ }\textbf {\bibinfo {volume} {98}},\ \bibinfo {pages} {062119} (\bibinfo {year} {2018})}\BibitemShut {NoStop}%
\bibitem [{\citenamefont {Wang}(2020)}]{wang_performance_2020}%
  \BibitemOpen
  \bibfield  {author} {\bibinfo {author} {\bibfnamefont {Q.}~\bibnamefont {Wang}},\ }\bibfield  {title} {\bibinfo {title} {Performance of quantum heat engines under the influence of long-range interactions},\ }\href {https://doi.org/10.1103/PhysRevE.102.012138} {\bibfield  {journal} {\bibinfo  {journal} {Physical Review E}\ }\textbf {\bibinfo {volume} {102}},\ \bibinfo {pages} {012138} (\bibinfo {year} {2020})}\BibitemShut {NoStop}%
\bibitem [{\citenamefont {S.}\ \emph {et~al.}(2025)\citenamefont {S.}, \citenamefont {Jaseem},\ and\ \citenamefont {Poletti}}]{s_quantum_2025}%
  \BibitemOpen
  \bibfield  {author} {\bibinfo {author} {\bibfnamefont {G.}~\bibnamefont {S.}}, \bibinfo {author} {\bibfnamefont {N.}~\bibnamefont {Jaseem}},\ and\ \bibinfo {author} {\bibfnamefont {D.}~\bibnamefont {Poletti}},\ }\bibfield  {title} {\bibinfo {title} {Quantum thermal machines and the emergence of different thermodynamic functioning regimes from finite coupling to a load},\ }\href {https://doi.org/10.1103/dr9b-5ryh} {\bibfield  {journal} {\bibinfo  {journal} {Physical Review A}\ }\textbf {\bibinfo {volume} {112}},\ \bibinfo {pages} {L050201} (\bibinfo {year} {2025})}\BibitemShut {NoStop}%
\bibitem [{\citenamefont {Bloch}\ \emph {et~al.}(2008)\citenamefont {Bloch}, \citenamefont {Dalibard},\ and\ \citenamefont {Zwerger}}]{bloch_many-body_2008}%
  \BibitemOpen
  \bibfield  {author} {\bibinfo {author} {\bibfnamefont {I.}~\bibnamefont {Bloch}}, \bibinfo {author} {\bibfnamefont {J.}~\bibnamefont {Dalibard}},\ and\ \bibinfo {author} {\bibfnamefont {W.}~\bibnamefont {Zwerger}},\ }\bibfield  {title} {\bibinfo {title} {Many-body physics with ultracold gases},\ }\href {https://doi.org/10.1103/RevModPhys.80.885} {\bibfield  {journal} {\bibinfo  {journal} {Reviews of Modern Physics}\ }\textbf {\bibinfo {volume} {80}},\ \bibinfo {pages} {885} (\bibinfo {year} {2008})}\BibitemShut {NoStop}%
\bibitem [{\citenamefont {Bloch}\ \emph {et~al.}(2012)\citenamefont {Bloch}, \citenamefont {Dalibard},\ and\ \citenamefont {Nascimbène}}]{bloch_quantum_2012}%
  \BibitemOpen
  \bibfield  {author} {\bibinfo {author} {\bibfnamefont {I.}~\bibnamefont {Bloch}}, \bibinfo {author} {\bibfnamefont {J.}~\bibnamefont {Dalibard}},\ and\ \bibinfo {author} {\bibfnamefont {S.}~\bibnamefont {Nascimbène}},\ }\bibfield  {title} {\bibinfo {title} {Quantum simulations with ultracold quantum gases},\ }\href {https://doi.org/10.1038/nphys2259} {\bibfield  {journal} {\bibinfo  {journal} {Nature Physics}\ }\textbf {\bibinfo {volume} {8}},\ \bibinfo {pages} {267} (\bibinfo {year} {2012})}\BibitemShut {NoStop}%
\bibitem [{\citenamefont {Bouton}\ \emph {et~al.}(2021)\citenamefont {Bouton}, \citenamefont {Nettersheim}, \citenamefont {Burgardt}, \citenamefont {Adam}, \citenamefont {Lutz},\ and\ \citenamefont {Widera}}]{bouton_quantum_2021}%
  \BibitemOpen
  \bibfield  {author} {\bibinfo {author} {\bibfnamefont {Q.}~\bibnamefont {Bouton}}, \bibinfo {author} {\bibfnamefont {J.}~\bibnamefont {Nettersheim}}, \bibinfo {author} {\bibfnamefont {S.}~\bibnamefont {Burgardt}}, \bibinfo {author} {\bibfnamefont {D.}~\bibnamefont {Adam}}, \bibinfo {author} {\bibfnamefont {E.}~\bibnamefont {Lutz}},\ and\ \bibinfo {author} {\bibfnamefont {A.}~\bibnamefont {Widera}},\ }\bibfield  {title} {\bibinfo {title} {A quantum heat engine driven by atomic collisions},\ }\href {https://doi.org/10.1038/s41467-021-22222-z} {\bibfield  {journal} {\bibinfo  {journal} {Nature Communications}\ }\textbf {\bibinfo {volume} {12}},\ \bibinfo {pages} {2063} (\bibinfo {year} {2021})}\BibitemShut {NoStop}%
\bibitem [{\citenamefont {Nettersheim}\ \emph {et~al.}(2022)\citenamefont {Nettersheim}, \citenamefont {Burgardt}, \citenamefont {Bouton}, \citenamefont {Adam}, \citenamefont {Lutz},\ and\ \citenamefont {Widera}}]{nettersheim_power_2022}%
  \BibitemOpen
  \bibfield  {author} {\bibinfo {author} {\bibfnamefont {J.}~\bibnamefont {Nettersheim}}, \bibinfo {author} {\bibfnamefont {S.}~\bibnamefont {Burgardt}}, \bibinfo {author} {\bibfnamefont {Q.}~\bibnamefont {Bouton}}, \bibinfo {author} {\bibfnamefont {D.}~\bibnamefont {Adam}}, \bibinfo {author} {\bibfnamefont {E.}~\bibnamefont {Lutz}},\ and\ \bibinfo {author} {\bibfnamefont {A.}~\bibnamefont {Widera}},\ }\bibfield  {title} {\bibinfo {title} {Power of a {Quasispin} {Quantum} {Otto} {Engine} at {Negative} {Effective} {Spin} {Temperature}},\ }\href {https://doi.org/10.1103/PRXQuantum.3.040334} {\bibfield  {journal} {\bibinfo  {journal} {PRX Quantum}\ }\textbf {\bibinfo {volume} {3}},\ \bibinfo {pages} {040334} (\bibinfo {year} {2022})}\BibitemShut {NoStop}%
\bibitem [{\citenamefont {Chin}\ \emph {et~al.}(2010)\citenamefont {Chin}, \citenamefont {Grimm}, \citenamefont {Julienne},\ and\ \citenamefont {Tiesinga}}]{chin_feshbach_2010}%
  \BibitemOpen
  \bibfield  {author} {\bibinfo {author} {\bibfnamefont {C.}~\bibnamefont {Chin}}, \bibinfo {author} {\bibfnamefont {R.}~\bibnamefont {Grimm}}, \bibinfo {author} {\bibfnamefont {P.}~\bibnamefont {Julienne}},\ and\ \bibinfo {author} {\bibfnamefont {E.}~\bibnamefont {Tiesinga}},\ }\bibfield  {title} {\bibinfo {title} {Feshbach resonances in ultracold gases},\ }\href {https://doi.org/10.1103/revmodphys.82.1225} {\bibfield  {journal} {\bibinfo  {journal} {Reviews of Modern Physics}\ }\textbf {\bibinfo {volume} {82}},\ \bibinfo {pages} {1225} (\bibinfo {year} {2010})}\BibitemShut {NoStop}%
\bibitem [{\citenamefont {Chen}\ \emph {et~al.}(2019)\citenamefont {Chen}, \citenamefont {Watanabe}, \citenamefont {Yu}, \citenamefont {Guan},\ and\ \citenamefont {Del~Campo}}]{chen_interaction-driven_2019}%
  \BibitemOpen
  \bibfield  {author} {\bibinfo {author} {\bibfnamefont {Y.-Y.}\ \bibnamefont {Chen}}, \bibinfo {author} {\bibfnamefont {G.}~\bibnamefont {Watanabe}}, \bibinfo {author} {\bibfnamefont {Y.-C.}\ \bibnamefont {Yu}}, \bibinfo {author} {\bibfnamefont {X.-W.}\ \bibnamefont {Guan}},\ and\ \bibinfo {author} {\bibfnamefont {A.}~\bibnamefont {Del~Campo}},\ }\bibfield  {title} {\bibinfo {title} {An interaction-driven many-particle quantum heat engine and its universal behavior},\ }\href {https://doi.org/10.1038/s41534-019-0204-5} {\bibfield  {journal} {\bibinfo  {journal} {npj Quantum Information}\ }\textbf {\bibinfo {volume} {5}},\ \bibinfo {pages} {88} (\bibinfo {year} {2019})}\BibitemShut {NoStop}%
\bibitem [{\citenamefont {Keller}\ \emph {et~al.}(2020)\citenamefont {Keller}, \citenamefont {Fogarty}, \citenamefont {Li},\ and\ \citenamefont {Busch}}]{keller_feshbach_2020}%
  \BibitemOpen
  \bibfield  {author} {\bibinfo {author} {\bibfnamefont {T.}~\bibnamefont {Keller}}, \bibinfo {author} {\bibfnamefont {T.}~\bibnamefont {Fogarty}}, \bibinfo {author} {\bibfnamefont {J.}~\bibnamefont {Li}},\ and\ \bibinfo {author} {\bibfnamefont {T.}~\bibnamefont {Busch}},\ }\bibfield  {title} {\bibinfo {title} {Feshbach engine in the {Thomas}-{Fermi} regime},\ }\href {https://doi.org/10.1103/PhysRevResearch.2.033335} {\bibfield  {journal} {\bibinfo  {journal} {Physical Review Research}\ }\textbf {\bibinfo {volume} {2}},\ \bibinfo {pages} {033335} (\bibinfo {year} {2020})}\BibitemShut {NoStop}%
\bibitem [{\citenamefont {Boubakour}\ \emph {et~al.}(2023)\citenamefont {Boubakour}, \citenamefont {Fogarty},\ and\ \citenamefont {Busch}}]{boubakour_interaction-enhanced_2023}%
  \BibitemOpen
  \bibfield  {author} {\bibinfo {author} {\bibfnamefont {M.}~\bibnamefont {Boubakour}}, \bibinfo {author} {\bibfnamefont {T.}~\bibnamefont {Fogarty}},\ and\ \bibinfo {author} {\bibfnamefont {T.}~\bibnamefont {Busch}},\ }\bibfield  {title} {\bibinfo {title} {Interaction-enhanced quantum heat engine},\ }\href {https://doi.org/10.1103/PhysRevResearch.5.013088} {\bibfield  {journal} {\bibinfo  {journal} {Physical Review Research}\ }\textbf {\bibinfo {volume} {5}},\ \bibinfo {pages} {013088} (\bibinfo {year} {2023})}\BibitemShut {NoStop}%
\bibitem [{\citenamefont {Nautiyal}\ \emph {et~al.}(2024)\citenamefont {Nautiyal}, \citenamefont {Watson},\ and\ \citenamefont {Kheruntsyan}}]{nautiyal_finite-time_2024}%
  \BibitemOpen
  \bibfield  {author} {\bibinfo {author} {\bibfnamefont {V.~V.}\ \bibnamefont {Nautiyal}}, \bibinfo {author} {\bibfnamefont {R.~S.}\ \bibnamefont {Watson}},\ and\ \bibinfo {author} {\bibfnamefont {K.~V.}\ \bibnamefont {Kheruntsyan}},\ }\bibfield  {title} {\bibinfo {title} {A finite-time quantum {Otto} engine with tunnel coupled one-dimensional {Bose} gases},\ }\href {https://doi.org/10.1088/1367-2630/ad57e5} {\bibfield  {journal} {\bibinfo  {journal} {New Journal of Physics}\ }\textbf {\bibinfo {volume} {26}},\ \bibinfo {pages} {063033} (\bibinfo {year} {2024})}\BibitemShut {NoStop}%
\bibitem [{\citenamefont {Watson}\ and\ \citenamefont {Kheruntsyan}(2025)}]{watson_quantum_2025}%
  \BibitemOpen
  \bibfield  {author} {\bibinfo {author} {\bibfnamefont {R.~S.}\ \bibnamefont {Watson}}\ and\ \bibinfo {author} {\bibfnamefont {K.}~\bibnamefont {Kheruntsyan}},\ }\bibfield  {title} {\bibinfo {title} {Quantum many-body thermal machines enabled by atom-atom correlations},\ }\href {https://doi.org/10.21468/SciPostPhys.18.6.190} {\bibfield  {journal} {\bibinfo  {journal} {SciPost Physics}\ }\textbf {\bibinfo {volume} {18}},\ \bibinfo {pages} {190} (\bibinfo {year} {2025})}\BibitemShut {NoStop}%
\bibitem [{\citenamefont {Wandhammer}\ \emph {et~al.}(2026)\citenamefont {Wandhammer}, \citenamefont {Hardel}, \citenamefont {Hervieux},\ and\ \citenamefont {Manfredi}}]{wandhammer_optimal_2026}%
  \BibitemOpen
  \bibfield  {author} {\bibinfo {author} {\bibfnamefont {A.}~\bibnamefont {Wandhammer}}, \bibinfo {author} {\bibfnamefont {V.}~\bibnamefont {Hardel}}, \bibinfo {author} {\bibfnamefont {P.-A.}\ \bibnamefont {Hervieux}},\ and\ \bibinfo {author} {\bibfnamefont {G.}~\bibnamefont {Manfredi}},\ }\bibfield  {title} {\bibinfo {title} {Optimal quantum {Feshbach} engines},\ }\href {https://doi.org/10.1103/ymmx-hskw} {\bibfield  {journal} {\bibinfo  {journal} {Physical Review A}\ }\textbf {\bibinfo {volume} {114}},\ \bibinfo {pages} {L040201} (\bibinfo {year} {2026})}\BibitemShut {NoStop}%
\bibitem [{\citenamefont {Simmons}\ \emph {et~al.}(2023)\citenamefont {Simmons}, \citenamefont {Sajjad}, \citenamefont {Keithley}, \citenamefont {Mas}, \citenamefont {Tanlimco}, \citenamefont {Nolasco-Martinez}, \citenamefont {Bai}, \citenamefont {Fredrickson},\ and\ \citenamefont {Weld}}]{simmons_thermodynamic_2023}%
  \BibitemOpen
  \bibfield  {author} {\bibinfo {author} {\bibfnamefont {E.~Q.}\ \bibnamefont {Simmons}}, \bibinfo {author} {\bibfnamefont {R.}~\bibnamefont {Sajjad}}, \bibinfo {author} {\bibfnamefont {K.}~\bibnamefont {Keithley}}, \bibinfo {author} {\bibfnamefont {H.}~\bibnamefont {Mas}}, \bibinfo {author} {\bibfnamefont {J.~L.}\ \bibnamefont {Tanlimco}}, \bibinfo {author} {\bibfnamefont {E.}~\bibnamefont {Nolasco-Martinez}}, \bibinfo {author} {\bibfnamefont {Y.}~\bibnamefont {Bai}}, \bibinfo {author} {\bibfnamefont {G.~H.}\ \bibnamefont {Fredrickson}},\ and\ \bibinfo {author} {\bibfnamefont {D.~M.}\ \bibnamefont {Weld}},\ }\bibfield  {title} {\bibinfo {title} {Thermodynamic engine with a quantum degenerate working fluid},\ }\href {https://doi.org/10.1103/PhysRevResearch.5.L042009} {\bibfield  {journal} {\bibinfo  {journal} {Physical Review Research}\ }\textbf {\bibinfo {volume} {5}},\ \bibinfo {pages} {L042009} (\bibinfo {year} {2023})}\BibitemShut {NoStop}%
\bibitem [{\citenamefont {Koch}\ \emph {et~al.}(2023)\citenamefont {Koch}, \citenamefont {Menon}, \citenamefont {Cuestas}, \citenamefont {Barbosa}, \citenamefont {Lutz}, \citenamefont {Fogarty}, \citenamefont {Busch},\ and\ \citenamefont {Widera}}]{koch_quantum_2023}%
  \BibitemOpen
  \bibfield  {author} {\bibinfo {author} {\bibfnamefont {J.}~\bibnamefont {Koch}}, \bibinfo {author} {\bibfnamefont {K.}~\bibnamefont {Menon}}, \bibinfo {author} {\bibfnamefont {E.}~\bibnamefont {Cuestas}}, \bibinfo {author} {\bibfnamefont {S.}~\bibnamefont {Barbosa}}, \bibinfo {author} {\bibfnamefont {E.}~\bibnamefont {Lutz}}, \bibinfo {author} {\bibfnamefont {T.}~\bibnamefont {Fogarty}}, \bibinfo {author} {\bibfnamefont {T.}~\bibnamefont {Busch}},\ and\ \bibinfo {author} {\bibfnamefont {A.}~\bibnamefont {Widera}},\ }\bibfield  {title} {\bibinfo {title} {A quantum engine in the {BEC}–{BCS} crossover},\ }\href {https://doi.org/10.1038/s41586-023-06469-8} {\bibfield  {journal} {\bibinfo  {journal} {Nature}\ }\textbf {\bibinfo {volume} {621}},\ \bibinfo {pages} {723} (\bibinfo {year} {2023})}\BibitemShut {NoStop}%
\bibitem [{\citenamefont {Menon}\ \emph {et~al.}(2025)\citenamefont {Menon}, \citenamefont {Busch},\ and\ \citenamefont {Fogarty}}]{menon_leveraging_2025}%
  \BibitemOpen
  \bibfield  {author} {\bibinfo {author} {\bibfnamefont {K.}~\bibnamefont {Menon}}, \bibinfo {author} {\bibfnamefont {T.}~\bibnamefont {Busch}},\ and\ \bibinfo {author} {\bibfnamefont {T.}~\bibnamefont {Fogarty}},\ }\bibfield  {title} {\bibinfo {title} {Leveraging quantum statistics to enhance heat engines},\ }\href {https://doi.org/10.1088/2058-9565/ae01d3} {\bibfield  {journal} {\bibinfo  {journal} {Quantum Science and Technology}\ }\textbf {\bibinfo {volume} {10}},\ \bibinfo {pages} {045039} (\bibinfo {year} {2025})}\BibitemShut {NoStop}%
\bibitem [{\citenamefont {Henriquez~Lira}\ \emph {et~al.}(2026)\citenamefont {Henriquez~Lira}, \citenamefont {Isaule}, \citenamefont {Groves}, \citenamefont {Peña}, \citenamefont {Vargas},\ and\ \citenamefont {Fogarty}}]{henriquez_lira_quantum_2026}%
  \BibitemOpen
  \bibfield  {author} {\bibinfo {author} {\bibfnamefont {S.}~\bibnamefont {Henriquez~Lira}}, \bibinfo {author} {\bibfnamefont {F.}~\bibnamefont {Isaule}}, \bibinfo {author} {\bibfnamefont {M.~H.}\ \bibnamefont {Groves}}, \bibinfo {author} {\bibfnamefont {F.~J.}\ \bibnamefont {Peña}}, \bibinfo {author} {\bibfnamefont {P.}~\bibnamefont {Vargas}},\ and\ \bibinfo {author} {\bibfnamefont {T.}~\bibnamefont {Fogarty}},\ }\href {https://doi.org/10.48550/arXiv.2609.37592} {\bibinfo {title} {Quantum {Statistical} {Thermal} {Engine} at the {BCS}-{BEC} crossover}} (\bibinfo {year} {2026}),\ \bibinfo {note} {arXiv:2609.37592}\BibitemShut {NoStop}%
\bibitem [{\citenamefont {Bloch}(2005)}]{bloch_ultracold_2005}%
  \BibitemOpen
  \bibfield  {author} {\bibinfo {author} {\bibfnamefont {I.}~\bibnamefont {Bloch}},\ }\bibfield  {title} {\bibinfo {title} {Ultracold quantum gases in optical lattices},\ }\href {https://doi.org/10.1038/nphys138} {\bibfield  {journal} {\bibinfo  {journal} {Nature Physics}\ }\textbf {\bibinfo {volume} {1}},\ \bibinfo {pages} {23} (\bibinfo {year} {2005})}\BibitemShut {NoStop}%
\bibitem [{\citenamefont {Lewenstein}\ \emph {et~al.}(2012)\citenamefont {Lewenstein}, \citenamefont {Sanpera},\ and\ \citenamefont {Ahufinger}}]{lewenstein_ultracold_2012}%
  \BibitemOpen
  \bibfield  {author} {\bibinfo {author} {\bibfnamefont {M.}~\bibnamefont {Lewenstein}}, \bibinfo {author} {\bibfnamefont {A.}~\bibnamefont {Sanpera}},\ and\ \bibinfo {author} {\bibfnamefont {V.}~\bibnamefont {Ahufinger}},\ }\href@noop {} {\emph {\bibinfo {title} {Ultracold atoms in optical lattices: simulating quantum many-body systems}}},\ \bibinfo {edition} {1st}\ ed.\ (\bibinfo  {publisher} {Oxford University Press},\ \bibinfo {address} {Oxford},\ \bibinfo {year} {2012})\BibitemShut {NoStop}%
\bibitem [{\citenamefont {Fogarty}\ and\ \citenamefont {Busch}(2021)}]{fogarty_many-body_2021}%
  \BibitemOpen
  \bibfield  {author} {\bibinfo {author} {\bibfnamefont {T.}~\bibnamefont {Fogarty}}\ and\ \bibinfo {author} {\bibfnamefont {T.}~\bibnamefont {Busch}},\ }\bibfield  {title} {\bibinfo {title} {A many-body heat engine at criticality},\ }\href {https://doi.org/10.1088/2058-9565/abbc63} {\bibfield  {journal} {\bibinfo  {journal} {Quantum Science and Technology}\ }\textbf {\bibinfo {volume} {6}},\ \bibinfo {pages} {015003} (\bibinfo {year} {2021})}\BibitemShut {NoStop}%
\bibitem [{\citenamefont {Peña}\ \emph {et~al.}(2026)\citenamefont {Peña}, \citenamefont {García-Zamora}, \citenamefont {Chiara}, \citenamefont {Flores}, \citenamefont {Henríquez}, \citenamefont {Myers}, \citenamefont {Barra},\ and\ \citenamefont {Vargas}}]{pena_bosonic_2026}%
  \BibitemOpen
  \bibfield  {author} {\bibinfo {author} {\bibfnamefont {F.~J.}\ \bibnamefont {Peña}}, \bibinfo {author} {\bibfnamefont {R.}~\bibnamefont {García-Zamora}}, \bibinfo {author} {\bibfnamefont {G.~D.}\ \bibnamefont {Chiara}}, \bibinfo {author} {\bibfnamefont {J.}~\bibnamefont {Flores}}, \bibinfo {author} {\bibfnamefont {S.}~\bibnamefont {Henríquez}}, \bibinfo {author} {\bibfnamefont {N.~M.}\ \bibnamefont {Myers}}, \bibinfo {author} {\bibfnamefont {F.}~\bibnamefont {Barra}},\ and\ \bibinfo {author} {\bibfnamefont {P.}~\bibnamefont {Vargas}},\ }\href {https://doi.org/10.48550/arXiv.2604.13185} {\bibinfo {title} {Bosonic {Working} {Media} in a {Frustrated} {Rhombi} {Chain}: {Otto} and {Stirling} {Cycles} from {Flat} {Bands}, {Caging}, and {Flux} {Control}}} (\bibinfo {year} {2026}),\ \bibinfo {note} {arXiv:2604.13185}\BibitemShut {NoStop}%
\bibitem [{\citenamefont {Lal~Bera}\ \emph {et~al.}(2026)\citenamefont {Lal~Bera}, \citenamefont {Kwan}, \citenamefont {Pérez}, \citenamefont {A.~García-March}, \citenamefont {Chhajlany}, \citenamefont {Grass}, \citenamefont {Lewenstein}, \citenamefont {Bhattacharya},\ and\ \citenamefont {Bhattacharjee}}]{lal_bera_hybrid_2026}%
  \BibitemOpen
  \bibfield  {author} {\bibinfo {author} {\bibfnamefont {M.}~\bibnamefont {Lal~Bera}}, \bibinfo {author} {\bibfnamefont {J.}~\bibnamefont {Kwan}}, \bibinfo {author} {\bibfnamefont {A.}~\bibnamefont {Pérez}}, \bibinfo {author} {\bibfnamefont {M.}~\bibnamefont {A.~García-March}}, \bibinfo {author} {\bibfnamefont {R.}~\bibnamefont {Chhajlany}}, \bibinfo {author} {\bibfnamefont {T.}~\bibnamefont {Grass}}, \bibinfo {author} {\bibfnamefont {M.}~\bibnamefont {Lewenstein}}, \bibinfo {author} {\bibfnamefont {U.}~\bibnamefont {Bhattacharya}},\ and\ \bibinfo {author} {\bibfnamefont {S.}~\bibnamefont {Bhattacharjee}},\ }\bibfield  {title} {\bibinfo {title} {A hybrid {Anyon}-otto thermal machine},\ }\href {https://doi.org/10.1038/s41534-026-01328-6} {\bibfield  {journal} {\bibinfo  {journal} {npj Quantum Information}\ }\textbf {\bibinfo {volume} {12}},\ \bibinfo {pages} {117} (\bibinfo {year} {2026})}\BibitemShut {NoStop}%
\bibitem [{\citenamefont {Fisher}\ \emph {et~al.}(1989)\citenamefont {Fisher}, \citenamefont {Weichman}, \citenamefont {Grinstein},\ and\ \citenamefont {Fisher}}]{fisher_boson_1989}%
  \BibitemOpen
  \bibfield  {author} {\bibinfo {author} {\bibfnamefont {M.~P.~A.}\ \bibnamefont {Fisher}}, \bibinfo {author} {\bibfnamefont {P.~B.}\ \bibnamefont {Weichman}}, \bibinfo {author} {\bibfnamefont {G.}~\bibnamefont {Grinstein}},\ and\ \bibinfo {author} {\bibfnamefont {D.~S.}\ \bibnamefont {Fisher}},\ }\bibfield  {title} {\bibinfo {title} {Boson localization and the superfluid-insulator transition},\ }\href {https://doi.org/10.1103/PhysRevB.40.546} {\bibfield  {journal} {\bibinfo  {journal} {Physical Review B}\ }\textbf {\bibinfo {volume} {40}},\ \bibinfo {pages} {546} (\bibinfo {year} {1989})}\BibitemShut {NoStop}%
\bibitem [{\citenamefont {Jaksch}\ \emph {et~al.}(1998)\citenamefont {Jaksch}, \citenamefont {Bruder}, \citenamefont {Cirac}, \citenamefont {Gardiner},\ and\ \citenamefont {Zoller}}]{jaksch_cold_1998}%
  \BibitemOpen
  \bibfield  {author} {\bibinfo {author} {\bibfnamefont {D.}~\bibnamefont {Jaksch}}, \bibinfo {author} {\bibfnamefont {C.}~\bibnamefont {Bruder}}, \bibinfo {author} {\bibfnamefont {J.~I.}\ \bibnamefont {Cirac}}, \bibinfo {author} {\bibfnamefont {C.~W.}\ \bibnamefont {Gardiner}},\ and\ \bibinfo {author} {\bibfnamefont {P.}~\bibnamefont {Zoller}},\ }\bibfield  {title} {\bibinfo {title} {Cold {Bosonic} {Atoms} in {Optical} {Lattices}},\ }\href {https://doi.org/10.1103/PhysRevLett.81.3108} {\bibfield  {journal} {\bibinfo  {journal} {Physical Review Letters}\ }\textbf {\bibinfo {volume} {81}},\ \bibinfo {pages} {3108} (\bibinfo {year} {1998})}\BibitemShut {NoStop}%
\bibitem [{\citenamefont {Di~Carli}\ \emph {et~al.}(2024)\citenamefont {Di~Carli}, \citenamefont {Parsonage}, \citenamefont {La~Rooij}, \citenamefont {Koehn}, \citenamefont {Ulm}, \citenamefont {Duncan}, \citenamefont {Daley}, \citenamefont {Haller},\ and\ \citenamefont {Kuhr}}]{di_carli_commensurate_2024}%
  \BibitemOpen
  \bibfield  {author} {\bibinfo {author} {\bibfnamefont {A.}~\bibnamefont {Di~Carli}}, \bibinfo {author} {\bibfnamefont {C.}~\bibnamefont {Parsonage}}, \bibinfo {author} {\bibfnamefont {A.}~\bibnamefont {La~Rooij}}, \bibinfo {author} {\bibfnamefont {L.}~\bibnamefont {Koehn}}, \bibinfo {author} {\bibfnamefont {C.}~\bibnamefont {Ulm}}, \bibinfo {author} {\bibfnamefont {C.~W.}\ \bibnamefont {Duncan}}, \bibinfo {author} {\bibfnamefont {A.~J.}\ \bibnamefont {Daley}}, \bibinfo {author} {\bibfnamefont {E.}~\bibnamefont {Haller}},\ and\ \bibinfo {author} {\bibfnamefont {S.}~\bibnamefont {Kuhr}},\ }\bibfield  {title} {\bibinfo {title} {Commensurate and incommensurate {1D} interacting quantum systems},\ }\href {https://doi.org/10.1038/s41467-023-44610-3} {\bibfield  {journal} {\bibinfo  {journal} {Nature Communications}\ }\textbf {\bibinfo {volume} {15}},\ \bibinfo {pages} {474} (\bibinfo {year} {2024})}\BibitemShut {NoStop}%
\bibitem [{\citenamefont {Zhang}\ and\ \citenamefont {Dong}(2010)}]{zhang_exact_2010}%
  \BibitemOpen
  \bibfield  {author} {\bibinfo {author} {\bibfnamefont {J.~M.}\ \bibnamefont {Zhang}}\ and\ \bibinfo {author} {\bibfnamefont {R.~X.}\ \bibnamefont {Dong}},\ }\bibfield  {title} {\bibinfo {title} {Exact diagonalization: the {Bose}–{Hubbard} model as an example},\ }\href {https://doi.org/10.1088/0143-0807/31/3/016} {\bibfield  {journal} {\bibinfo  {journal} {European Journal of Physics}\ }\textbf {\bibinfo {volume} {31}},\ \bibinfo {pages} {591} (\bibinfo {year} {2010})}\BibitemShut {NoStop}%
\bibitem [{\citenamefont {Raventós}\ \emph {et~al.}(2017)\citenamefont {Raventós}, \citenamefont {Graß}, \citenamefont {Lewenstein},\ and\ \citenamefont {Juliá-Díaz}}]{raventos_cold_2017}%
  \BibitemOpen
  \bibfield  {author} {\bibinfo {author} {\bibfnamefont {D.}~\bibnamefont {Raventós}}, \bibinfo {author} {\bibfnamefont {T.}~\bibnamefont {Graß}}, \bibinfo {author} {\bibfnamefont {M.}~\bibnamefont {Lewenstein}},\ and\ \bibinfo {author} {\bibfnamefont {B.}~\bibnamefont {Juliá-Díaz}},\ }\bibfield  {title} {\bibinfo {title} {Cold bosons in optical lattices: a tutorial for exact diagonalization},\ }\href {https://doi.org/10.1088/1361-6455/aa68b1} {\bibfield  {journal} {\bibinfo  {journal} {Journal of Physics B: Atomic, Molecular and Optical Physics}\ }\textbf {\bibinfo {volume} {50}},\ \bibinfo {pages} {113001} (\bibinfo {year} {2017})}\BibitemShut {NoStop}%
\bibitem [{\citenamefont {Kühner}\ and\ \citenamefont {Monien}(1998)}]{kuhner_phases_1998}%
  \BibitemOpen
  \bibfield  {author} {\bibinfo {author} {\bibfnamefont {T.~D.}\ \bibnamefont {Kühner}}\ and\ \bibinfo {author} {\bibfnamefont {H.}~\bibnamefont {Monien}},\ }\bibfield  {title} {\bibinfo {title} {Phases of the one-dimensional {Bose}-{Hubbard} model},\ }\href {https://doi.org/10.1103/PhysRevB.58.R14741} {\bibfield  {journal} {\bibinfo  {journal} {Physical Review B}\ }\textbf {\bibinfo {volume} {58}},\ \bibinfo {pages} {R14741} (\bibinfo {year} {1998})}\BibitemShut {NoStop}%
\bibitem [{\citenamefont {Wessel}\ \emph {et~al.}(2004)\citenamefont {Wessel}, \citenamefont {Alet}, \citenamefont {Troyer},\ and\ \citenamefont {Batrouni}}]{wessel_quantum_2004}%
  \BibitemOpen
  \bibfield  {author} {\bibinfo {author} {\bibfnamefont {S.}~\bibnamefont {Wessel}}, \bibinfo {author} {\bibfnamefont {F.}~\bibnamefont {Alet}}, \bibinfo {author} {\bibfnamefont {M.}~\bibnamefont {Troyer}},\ and\ \bibinfo {author} {\bibfnamefont {G.~G.}\ \bibnamefont {Batrouni}},\ }\bibfield  {title} {\bibinfo {title} {Quantum {Monte} {Carlo} simulations of confined bosonic atoms in optical lattices},\ }\href {https://doi.org/10.1103/PhysRevA.70.053615} {\bibfield  {journal} {\bibinfo  {journal} {Physical Review A}\ }\textbf {\bibinfo {volume} {70}},\ \bibinfo {pages} {053615} (\bibinfo {year} {2004})}\BibitemShut {NoStop}%
\bibitem [{\citenamefont {Verstraete}\ \emph {et~al.}(2004)\citenamefont {Verstraete}, \citenamefont {García-Ripoll},\ and\ \citenamefont {Cirac}}]{verstraete_matrix_2004}%
  \BibitemOpen
  \bibfield  {author} {\bibinfo {author} {\bibfnamefont {F.}~\bibnamefont {Verstraete}}, \bibinfo {author} {\bibfnamefont {J.~J.}\ \bibnamefont {García-Ripoll}},\ and\ \bibinfo {author} {\bibfnamefont {J.~I.}\ \bibnamefont {Cirac}},\ }\bibfield  {title} {\bibinfo {title} {Matrix {Product} {Density} {Operators}: {Simulation} of {Finite}-{Temperature} and {Dissipative} {Systems}},\ }\href {https://doi.org/10.1103/PhysRevLett.93.207204} {\bibfield  {journal} {\bibinfo  {journal} {Physical Review Letters}\ }\textbf {\bibinfo {volume} {93}},\ \bibinfo {pages} {207204} (\bibinfo {year} {2004})}\BibitemShut {NoStop}%
\bibitem [{\citenamefont {Feiguin}\ and\ \citenamefont {White}(2005)}]{feiguin_finite-temperature_2005}%
  \BibitemOpen
  \bibfield  {author} {\bibinfo {author} {\bibfnamefont {A.~E.}\ \bibnamefont {Feiguin}}\ and\ \bibinfo {author} {\bibfnamefont {S.~R.}\ \bibnamefont {White}},\ }\bibfield  {title} {\bibinfo {title} {Finite-temperature density matrix renormalization using an enlarged {Hilbert} space},\ }\href {https://doi.org/10.1103/PhysRevB.72.220401} {\bibfield  {journal} {\bibinfo  {journal} {Physical Review B}\ }\textbf {\bibinfo {volume} {72}},\ \bibinfo {pages} {220401} (\bibinfo {year} {2005})}\BibitemShut {NoStop}%
\end{thebibliography}%

\end{document}